\pdfoutput=1
\documentclass{elsarticle}
\usepackage{amsmath}    
\usepackage{graphicx}   
\usepackage{listing}
\usepackage{color}
\usepackage{url}

\definecolor{gray}{rgb}{0.7,0.7,0.7}

\begin{document}
\title{AiiDA: Automated Interactive Infrastructure and Database for Computational Science}

\author[epfl]{Giovanni Pizzi\fnref{fn1}}
 \ead{giovanni.pizzi@epfl.ch}   
\author[epfl]{Andrea Cepellotti\fnref{fn1}}
 \ead{andrea.cepellotti@epfl.ch}   
\author[epfl]{Riccardo Sabatini}
\author[epfl]{Nicola Marzari}
\author[bosch]{Boris Kozinsky}
\address[epfl]{Theory and Simulation of Materials (THEOS), and National Centre for Computational Design and Discovery of Novel Materials (MARVEL), \'Ecole Polytechnique F\'ed\'erale de Lausanne, 1015 Lausanne, Switzerland}
\address[bosch]{Research and Technology Center, Robert Bosch LLC, Cambridge, Massachusetts 02139, USA}
\fntext[fn1]{These two authors equally contributed to the work.}
\begin{abstract}
\footnotesize
Computational science has seen in the last decades a spectacular rise
in the scope, breadth, and depth of its efforts. Notwithstanding this
prevalence and impact, it is often still performed using the renaissance model of individual artisans gathered in a workshop, under 
the guidance of an established practitioner. Great benefits could follow 
instead from adopting concepts and tools coming from computer science to manage, preserve, and share these computational efforts. We illustrate here our paradigm sustaining such vision, based around the four pillars of 
Automation, Data, Environment, and Sharing. We then discuss 
its implementation in the open-source AiiDA platform
(http://www.aiida.net), that has been tuned first to the demands of
computational materials science.
AiiDA's design is based on directed acyclic graphs to track the
provenance of data and calculations, and ensure preservation and
searchability. Remote computational resources are managed
transparently, and automation is coupled with data storage to ensure
reproducibility. Last, complex sequences of calculations can be encoded into
scientific workflows. We believe that AiiDA's design and its sharing
capabilities will encourage the creation of social ecosystems to
disseminate codes, data, and scientific workflows. 
\end{abstract}

\maketitle

\section{Introduction}
Computational science has now emerged as a new paradigm straddling
experiments and theory. While in the early days of computer
simulations only a small number of expensive simulations could be performed,
nowadays the capacity of computer simulations
calls for the development of concepts
and tools to organize research efforts with the ideas and practices of 
computer science.

In this paper, we take as a case study computational materials
science, given that the accuracy and predictive power of the
current ``quantum engines'' (computational
codes performing quantum-mechanical simulations) have made
these widespread in science and
technology, to the point that nowadays they are routinely used
in industry and academia to understand, predict, and design 
properties of complex materials and devices.
This success is also demonstrated by the fact that 12 out of the
top-100 most cited papers in the entire scientific literature
worldwide deal with quantum simulations using density-functional
theory~\cite{VanNoorden:2014}.
The availability of robust codes based on these
accurate physical models, the sustained increase in high-performance
computational (HPC) capacity, and the appearance of curated materials
databases has paved the way for the emergence of the model of
materials design and discovery via high-throughput computations (see Refs.~\cite{Franceschetti:1999,Johannesson:2002,Curtarolo:2003}, and
Ref.~\cite{Curtarolo:2013} for the literature cited therein).
In some realizations of this paradigm many different materials
(taken from experimental or computational databases)
and their properties are screened for optimal performance.
This type of work requires to automate the computational engines
in order to run thousands of simulations or more.

The challenge of automating and managing the simulations and the 
resulting data creates the need for a dedicated software infrastructure.
However, a practical implementation of a flexible and general tool
is challenging, because it must address two contrasting
requirements. On one hand
it must be flexible enough to support different tasks.
On the other hand, it must require minimal effort to be used.
While addressing these opposite requirements,
two additional challenges should be tackled:
ensuring reproducibility of simulations, and encouraging the creation 
of a community to share and cross-validate results.
Regarding the former, even nowadays many 
scientific computational papers do not provide all the 
details needed to reproduce the results.
As for the latter, a common suggestion to encourage sharing is to create
unified central databanks of computed results, where users
can contribute their data~\cite{Villars:2004,Zarkevich:2006,Silveira:2008,Yuan:2010,Jain:2011,Adams:2011,Curtarolo:2012,Landis:2012,Saal:2013,eudat,nomad}.
However, the bulk of these repositories comes often
from the group initiating the database.
One reason could be that researchers do not
perceive the benefits of sharing data in a competitive
academic or commercial environment.
The main obstacle, though, is that even researchers who are willing to 
contribute do not have the ability or the time to easily collect
their own data, convert these to another format, filter and upload them.

With these considerations in mind, we developed AiiDA,
an Automated Interactive Infrastructure and Database 
for computational science. 
Using AiiDA, the users can access transparently
both local and remote computer resources. The platform is easy to
use thanks to a high-level python scripting interface and can support
different codes by means of a plugin interface.
A central goal of AiiDA is the full reproducibility of calculations
and of the resulting data chain, that we  
obtain by a tight coupling of storage and workflow automation.
Data analysis and querying of heterogeneous results and of the
provenance relationships are made possible and effective by a database
design based on directed acyclic graphs and
targeted towards data management for high-throughput simulations.
Sharing of scientific knowledge is addressed by making it easy to
setup and manage local repositories driven by the interests of a given
group, but providing tools to seamlessly share not only the data
itself, but also the full scientific workflows used to generate the results.

In this paper, we first discuss in more detail the general
requirements that any infrastructure should have
to create, manage, analyze and share data and simulations.
These requirements are summarized in the
four pillars of the ADES model (Automation, Data, 
Environment, and Sharing). We then describe in detail how these
have been addressed by the current open-source implementation of AiiDA,
starting from Sec.~\ref{sec:automation} (one section per pillar).

\section{\label{sec:requirements}The ADES model for computational science}

\begin{figure}[t]
\includegraphics[width=\linewidth]{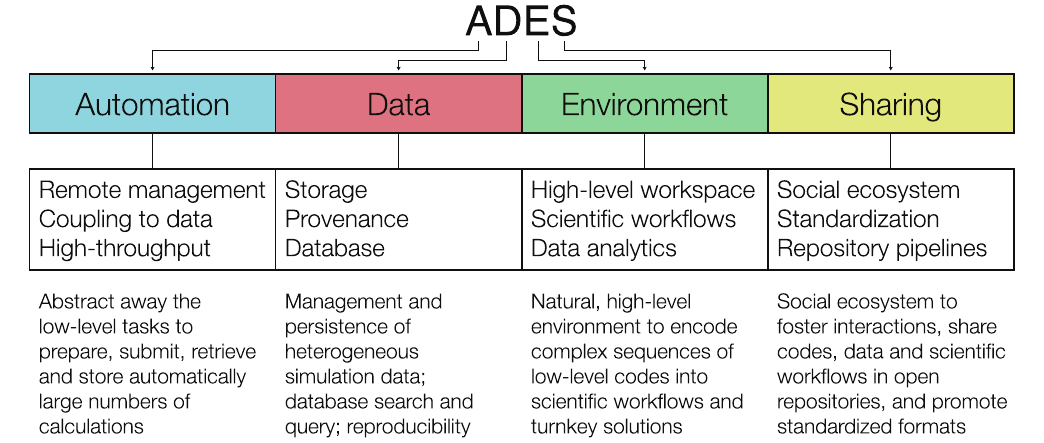}

\caption{\label{fig:infrastructurepillars}The four pillars of the proposed infrastructure
  for computational science. At the lower level,
  an automation framework and an efficient data management solution are 
  needed. At the user level,
  a high-level environment is coupled with a social ecosystem 
  to stimulate the sharing of codes, data and workflows.}
\end{figure}

The aim of this section is to introduce and illustrate the ADES
model (see Fig.~\ref{fig:infrastructurepillars}), in
order to motivate our design choices for AiiDA and describe the platform
requirements.

\vspace{6pt}
The first pillar, \textbf{Automation}, responds to the needs of abstracting away
the low-level tasks to prepare, submit, retrieve and store 
large numbers of calculations. It can be subdivided into
the following main items:

\begin{itemize}
\item \textbf{Remote management} 
Large computations are typically prepared on a user's workstation and executed on HPC clusters. 
The different steps of job preparation, submission, status check, and
results retrieval are repetitive and independent of the specific
simulation tool. Therefore, remote management tasks can 
be abstracted into an Application Programming Interface (API) and 
automated. Different communication and scheduler
implementations can be supported
by plugins, all adopting the same API, as we discuss in
Sec.~\ref{sec:transportandscheduler}.

\item \textbf{Coupling to data} 
Full reproducibility of calculations requires
a \emph{tight coupling of automation and storage}.
Decoupling these two aspects
leaves the researcher with the error-prone task of 
manually uploading calculation inputs and outputs 
to a suitable repository, with the additional
risk of providing  incomplete information. 
Instead, if the repository is populated first 
by the user with all the information needed to run the simulation,
the process of creating the input files and
running the calculation can be automated and
easily repeated. 
The resulting repositories are therefore necessarily consistent;
moreover, almost no user intervention is required to create
direct pipelines to shared repositories, as the data is already
stored coherently.

\item \textbf{High-throughput}
The main advantage of automation is obtained in situations when screening or parameter sweeps are required, 
involving thousands of calculations or more.
Running and managing them one by one is 
not feasible.
Having a high-level automation framework opens
the possibility to run multiple calculations simultaneously,
analyze and filter the results.
The infrastructure must be able to deal with 
potential errors arising during the computations,
trying to automatically recognize and remedy these whenever possible.

\end{itemize}

The second pillar, \textbf{Data}, concerns the management of the data
produced by the simulations and covers the following three core areas:
\begin{itemize}

\item \textbf{Storage} 
HPC calculations produce a large amount of heterogeneous data.
Files containing input parameters and final
results need to be automatically and permanently stored for future
reference and analysis. On the other hand, much of the data is required only 
temporarily (e.g., for check-pointing) and can be discarded at
the end of the simulation. 
Therefore, a code-dependent file-storage policy (optionally
customizable by the user) must be adopted to categorize each output file. 
Anyhow, the existence of intermediate files should be recorded,
so that the logical flow of calculations is persisted even
when restart files are deleted. If the platform ensures
reproducibility of calculations, it is straightforward to
regenerate the intermediate files, if needed.
It is also important to store information on the codes that generated
the data. If ultimate reproducibility is
needed, one could envision to store reference virtual machines
or Docker~\cite{docker} images with the code executables.

\item \textbf{Provenance}
To achieve reproducibility, the platform needs to store and represent the
calculations that are executed, together with their input data.
An effective data model, though, should not only put emphasis on 
calculations and data, but also keep track of the causal 
relationships between them, i.e.,
the full provenance of the results. For instance, 
a final relaxed crystal structure is of limited use
without knowing how it was obtained. 
The natural data structure to represent the network of relations
between data and calculations is a directed acyclic graph,
as we will motivate in greater detail in Sec. 4.1.

\item \textbf{Database} Today's typical computational work environment
consists of a multitude of files with arbitrary directory
structures, naming schemes and lacking documentation. In practice, it is
hard to understand and use the information (even by the author
after some time) and to retrieve a specific
calculation when many are stored.
A database can help in organizing results and querying
them.
The implementation of the data model
discussed above, based on directed acyclic graphs,
must not be restricted to a specific application, but has to accommodate
heterogeneous data.
It must be possible to efficiently query
any attribute (number, string, list, dictionary, \ldots) associated 
to a graph node. Queries that traverse
the graph to assess causal relationships between nodes
must also be possible.
A graph database backend is not required if the requirements above 
are satisfied. For instance, AiiDA's backend is a relational database
with a transitive-closure table for efficient graph-traversal
(see Secs.~\ref{sec:db} and \ref{sec:graphdb}).
\end{itemize}

The first two pillars described above address mainly low-level
functionalities. The next two pillars deal instead with user-oriented features.
In particular, the pillar \textbf{Environment} focuses on
creating a natural environment for computational
science, and involves the following aspects:

\begin{itemize}
\item \textbf{High-level workspace} 
As the researcher's objective is to make new discoveries and
not to learn a new code, the infrastructure should be
flexible and straightforward to use.
For instance, while databases offer many advantages in data-driven
computational science, few scientists are expert in their administration.
For this reason, the intricacies of database management and connections
must be hidden by the an API abstraction layer.
Furthermore, by adopting a widespread high-level programming language (such
as Python) one can benefit of mature tools for inserting and 
retrieving data from databases~\cite{django,sqlalchemy,rubyonrails}. 
The infrastructure must also be modular: a core providing
common low-level functionalities, and customizable plugins
to support different codes.

\item \textbf{Scientific workflows} 
  Much of the scientific knowledge does not merely lie in the final data,
  but in the description of the process, i.e., the ``scientific workflow''
  used to obtain them. 
  If these processes can be encoded, then they can be
  reused to compute similar quantities in different contexts.
  A workflow specifies a dependency tree between calculation steps, that
  may not be defined at the start, but depend
  on intermediate results (e.g., an iterative convergence with an
  unpredictable number of iterations). Therefore, the infrastructure
  should automatically generate dependent calculations only when
  their inputs are  available from earlier steps,
  evaluating dependencies at run-time.
  The integration of the scientific workflows with the other 
  infrastructure pillars helps the users to focus
  on the workflow logic rather than
  on the details of the remote management. 
  As an additional benefit, the automatic storage of the
  provenance during execution provides an implicit
  documentation of the logic behind the results.

\item \textbf{Data analytics}
Application-driven research has the necessity of using
dozens of different tools and approximations. 
Nevertheless, results obtained with different codes often
require the same post-processing or visualization algorithms.
These data types (e.g., crystal structures or
band structures) should be stored in the same common format.
The infrastructure can then either provide data analytics
capabilities to perform operations on them, or
even better facilitate the adoption of existing libraries.
This result can be achieved by providing interfaces
to external tools for data processing and
analytics (e.g.~\cite{Bahn:2002,Ong:2013} for crystal structures),
regardless of the specific simulation code used to generate the data.
\end{itemize}

The fourth pillar, \textbf{Sharing}, envisions the creation of a social
ecosystem to foster interaction between scientists, in particular for 
sharing data, results and scientific workflows:

\begin{itemize}
\item \textbf{Social ecosystem}
The envisioned framework should be an enabling technology to create
a {\it social ecosystem} in computational research.
Data access policies must be considered with great care.
Researchers prefer at times to keep their data private
(while protecting information in pending patents or unpublished data), 
but sharing with collaborators or on a public repository
should occur with minimal effort, when desired.
Beside data sharing, 
a standardized plugin interface should be provided.
Plugin repositories can be set up, to which users can contribute
to share workflows, handlers for new data formats, or support 
for new simulations codes.
By this mechanism, scientists will be able to engage in {\it social
  computing}, 
parallel to the developments in the mobile app and web ecosystems.

\item \textbf{Standardization} In order to 
  facilitate data exchange, standard formats should be agreed upon 
  and adopted for data sharing (e.g.~\cite{Murray-Rust:2003}). 
  Even when multiple standards exist, a hub-and-spoke configuration
  can be envisaged, where each new code has the task to provide the data
  in an established format. 
  On the other hand, it is important that suitable ontologies are defined
  (i.e., simplifying, the names and physical units of the quantities
  to store in a given repository, together with their meaning).
  Ontologies are field-specific and their definition must be
  community-driven (an example of an ongoing 
  effort is the TCOD~\cite{tcod} database).
  The infrastructure can be useful in this respect both as an
  inspiration for the ontology, and as a testing environment 
  containing a set of simulated use cases.

\item \textbf{Repository pipelines} As more repositories emerge,
it is important to develop the ability to import or export data directly, either
through REST interfaces or via suitably defined protocols.
If formats and ontologies are established, 
the platform must simply convert the data and its provenance in the
specified format. Contributing to external databases becomes 
straightforward and the platform becomes a facilitator for the
creation of shared repositories.
\end{itemize}

\section{\label{sec:infrastructure}The AiiDA infrastructure}
The ADES model described in the previous section
aims at defining an integrated infrastructure for
automating, storing, managing and sharing simulations and their results.
Until now, we discussed the model at an
abstract level, so as to highlight the generality of the requirements.
In order to 
provide researchers with an effective tool to manage their efforts,
we developed a Python infrastructure (``AiiDA'', http://www.aiida.net) that
is distributed open-source.
In the following, we describe the implementation details
of AiiDA, with particular emphasis on how the requirements of
Sec.~\ref{sec:requirements} have been met.

\begin{figure}[t]
\centering\includegraphics[width=12cm]{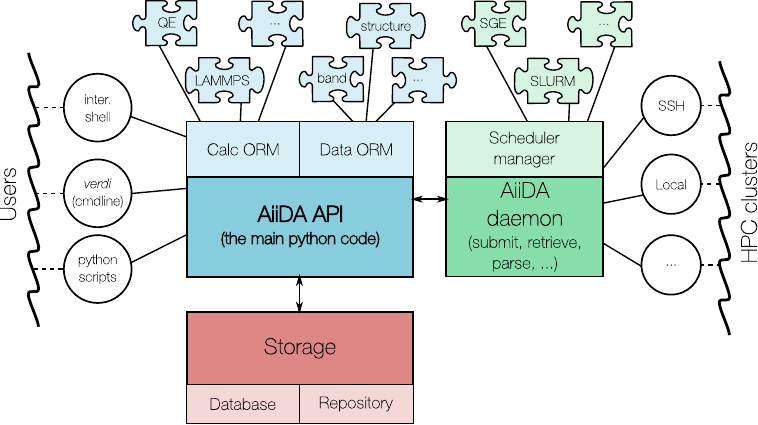}
\caption{\label{fig:infrastructure}The main components of the AiiDA
infrastructure and their interactions. The core AiiDA component is the API,
whose ORM represents stored objects as
python classes.
AiiDA supports any computational code and data
type via a plugin interface.
The AiiDA daemon is a background process that takes
care of most automated operations such as job submission,
scheduler state check, file retrieval and parsing. 
It interacts with the remote clusters via different
channels (local, ssh, \ldots) using the appropriate scheduler
plugins.}
\end{figure}

We start by outlining the architecture of AiiDA,
schematically represented in Fig.~\ref{fig:infrastructure}.
AiiDA has been designed as an intermediate layer between the user and the
HPC resources, where automation is achieved by \emph{abstraction}.
 
The core of the code is represented by the AiiDA API, a set of 
Python classes that expose to the user 
an intuitive interface to interact with the main AiiDA
objects --- calculations, codes and data --- hiding the inhomogeneities
of different supercomputers or data storage solutions.
The key component of
the API is the Object--Relational Mapper (ORM), 
a layer that maps AiiDA storage objects into python
classes. Using the ORM, these objects can be created, modified and queried
via a high-level interface which is agnostic of the detailed storage solution or of the SQL query language. The details of the storage, composed of both a relational database
and a file repository, are discussed in Sec.~\ref{sec:data}.

The user interacts with AiiDA in different ways:
using the command line tool \texttt{verdi}, via the interactive
python shell, or directly through python scripts (more details in 
Sec.~\ref{sec:verdi}).
Most components are designed with a plugin architecture (Sec.~\ref{sec:plugins}). 
Examples of features that can be extended with
new plugins include the support of new simulation codes,
management of new data types, and connection to remote
computers using different job schedulers.

\section{\label{sec:automation}Automation in AiiDA}

\subsection{\label{sec:daemon}The AiiDA daemon}
The daemon is one important building block of AiiDA:
it is a process that runs in the
background and handles the interaction with HPC clusters
(selecting the appropriate plugins for the 
communication channels --- like SSH --- or for the different
job schedulers, see Sec.~\ref{sec:transportandscheduler} below) and
takes care of all automation tasks.
Once the daemon is started, it runs in the background,
so that users can even log out from their accounts without
stopping AiiDA. Internally, it uses \texttt{celery}~\cite{celery}
and \texttt{supervisor}~\cite{supervisor} to manage asynchronous
tasks.

The fundamental role of the daemon is to manage the life cycle
of single calculations. The management operations are implemented
in the \texttt{aiida.execmanager}
module and consists in three main tasks: 1) submission of a new job to a remote computer, 
2) verification of the remote job scheduler state,
and 3) retrieval and parsing of the results after a job completion.
These steps are run independently.
If several calculations are running on the same machine,
they are grouped in order to open only one remote
connection and avoid to overload the remote cluster.

The user can follow the evolution of a calculation without
connecting directly to the remote machine by checking the 
\texttt{state} of a calculation, an attribute that is
stored in the database and is constantly updated by the daemon.
In particular, every calculation is initialized in 
a \texttt{NEW} state. A call to the \texttt{calc.submit()}
method brings \texttt{calc} to the \texttt{TOSUBMIT} state.
As soon as the daemon discovers a new calculation with
this state, it performs all the necessary
operations to submit
the calculation and then sets the state to 
\texttt{WITHSCHEDULER}.
Periodically, the daemon checks the remote scheduler state of
\texttt{WITHSCHEDULER} calculations and,
at job completion, the relevant files
are automatically retrieved, parsed and
saved in the AiiDA storage.
Finally, the state of the calculation is set to
\texttt{FINISHED}, or \texttt{FAILED} if the parser
detects that the calculation did not complete correctly.
Beside the aforementioned states, other transition
states exist (\texttt{SUBMITTING},
\texttt{RETRIEVING}, \texttt{PARSING}) as well as states to identify
failures occurred in specific states (\texttt{SUBMISSIONFAILED},
\texttt{RETRIEVALFAILED} and
\texttt{PARSINGFAILED}).

\subsection{\label{sec:transportandscheduler}Transports and schedulers}
As discussed in the ``Remote management'' section of the ``Automation'' pillar,
in AiiDA we define an abstract API layer with methods
to connect and communicate with remote computers 
and to interact with the schedulers. Thanks to this API,
the internal AiiDA code and the user interface are independent
of the type of connection protocol and scheduler that are actually used. 

The generic job attributes valid for any scheduler
(wall clock time, maximum required memory,
name of the output files, ...) are stored in a common format.
For what concerns schedulers, early work in the specification of a
middleware API has been done in the Open Grid
Forum~\cite{opengridforum} with, e.g., the
DRMAA~\cite{Troger:2007} and the SAGA APIs, and similar efforts have
been done by the UNICORE~\cite{unicore} and gc3pie~\cite{gc3pie}
projects. In AiiDA, we have taken inspiration from these efforts.
We provide appropriate plugins to convert the abstract information to the specific
headers to be written at the top of the scheduler submission file. Moreover,
the plugins provide methods that specify how
to submit a new job or how to retrieve the job state 
(running, queued, \ldots).
Plugins for the most common job schedulers (Torque~\cite{torque},
PBS Professional~\cite{pbspro}, SLURM~\cite{slurm},
SGE or its forks~\cite{opengridscheduler}) are already provided
with AiiDA.

The scheduler plugins and the daemon, then, rely
on the transport component to perform the necessary remote operations (file
copy and transfer, command execution, \ldots).
Also in this case, we have defined an abstract API specifying the
standard commands that should be available on any transport channel
(connection open and close, file upload and download, file list,
command execution, \ldots). Plugins define the specific implementation.
With AiiDA, we provide a \texttt{local} transport plugin, to be used if
AiiDA is installed on the same cluster on which calculations will be
executed. This plugin performs directly command execution and file
copy using the \texttt{os} and \texttt{shutil} Python
modules. We also provide a \texttt{ssh} transport plugin
to connect to remote machines using an encrypted and authenticated 
SSH channel, and SFTP for file transfer. In this case,
AiiDA relies on \texttt{paramiko}~\cite{paramiko} for the Python
implementation of the SSH and SFTP protocols.

The appropriate plugins to be used for each of the configured
computers are specified only once, when user configures for the first time
a new remote computer in AiiDA.

 \section{\label{sec:data}Data in AiiDA: database, storage and provenance}
\subsection{\label{sec:openprovenance}The data model in AiiDA}
The core concept of the AiiDA data
model, partially inspired by the Open Provenance Model~\cite{Moreau:2011},
is that any calculation acts as a function (with the meaning this word
has in mathematics or in a computer language), performing
some manipulation on a set of input data to produce new data as output.

\begin{figure}[t]
\centering\includegraphics[width=\linewidth]{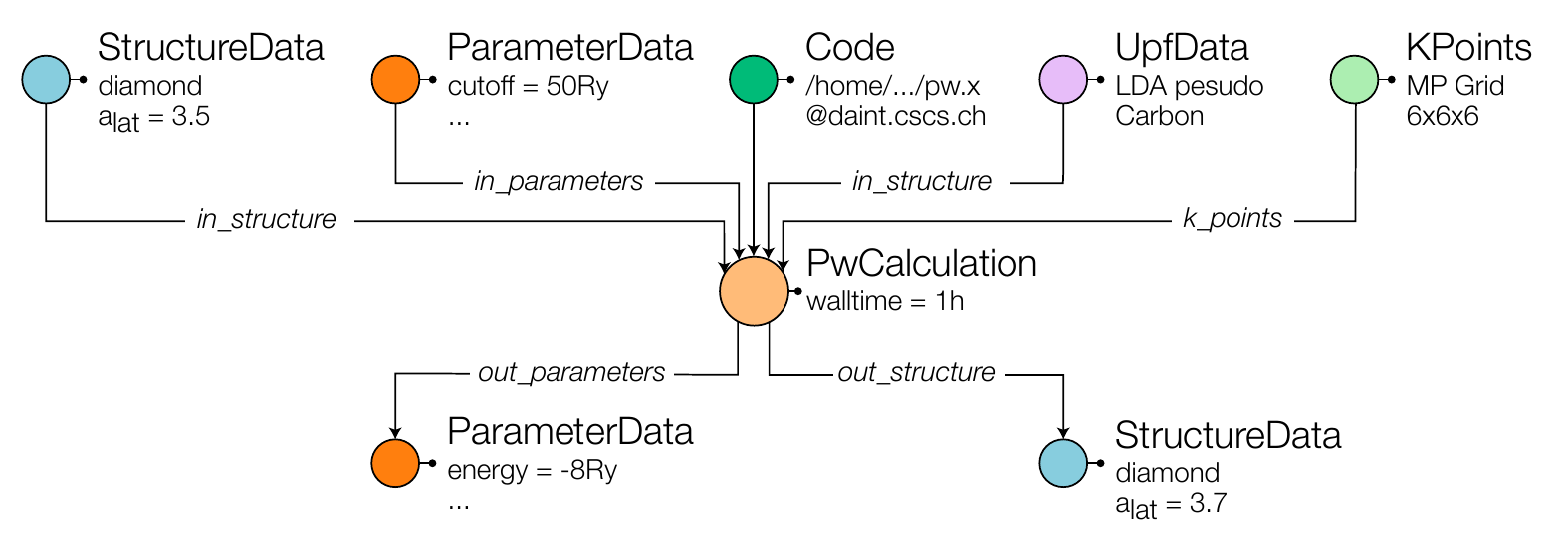}
\caption{\label{fig:simpleprovenance}A simple example of how a
calculation, the executable code
and input/output data are represented as nodes in a graph.
Labeled links between nodes represent logical relationships: either 
inputs or outputs. The code in input of a calculation
represents the executable that is launched.
In this example, a Quantum ESPRESSO code is used to
relax a diamond crystal, using further parameters as
input nodes (cutoffs, a mesh of $k-$points, a pseudopotential in UPF format for carbon, \ldots).
In output, two nodes are produced: a list of result parameters
(e.g., the total energy) and an output relaxed structure.
This node can in turn be used as input of new
calculations.}
\end{figure}

We thus represent each fundamental object, \texttt{Calculation}
and \texttt{Data}, as a node in a graph.
These nodes can be connected together with directional
and labeled links to 
represent input and output data of a calculation. Direct links
between \texttt{Data} nodes are not allowed: any operation (even a simple
copy) converting data objects to other data objects is a function 
and must thus be represented by an intermediate
\texttt{Calculation} node. 
We define for convenience a third fundamental object,
the \texttt{Code}, representing the executable file that is
run on the HPC resource. Each \texttt{Calculation} has therefore
a set of \texttt{Data} nodes and a \texttt{Code} node as input
(Fig.~\ref{fig:simpleprovenance}).
As the output \texttt{Data} nodes can in turn be used as input
of new calculations, we are effectively modeling a Directed Acyclic Graph (DAG) representing
the chain of relationships between the initial data (e.g., a
crystal structure from an experimental database) and the final
results (e.g., a luminescence spectrum) through all
the intermediate steps that are required to obtain the final result:
the provenance information of the data is therefore saved. (The graph is acyclic because links represent a causal connection, and therefore a loop is not allowed.)

\subsection{\label{sec:db}The AiiDA database}

Given that AiiDA represents data in terms of DAGs,
we need to choose an efficient way to save them on disk. 
The objects we need to store are the nodes and the links 
between them. Each node needs to contain all the information describing it,
such as lists of input flags, sets of parameters, list of coordinates,
possibly some files, etc. Therefore, the actual implementation must
support the storage of arbitrary lists of
files, and of attributes in the form \texttt{key=value} (of different types: 
strings, numbers, lists, \ldots) associated to each node.
One simple solution could consist in storing one file per node,
containing all node attributes in a suitable format, and then store
all the links in another file.
However, this storage type is clearly not efficient for querying, 
because in the absence of a suitable
indexing system every search requires disk access to each file.
A database, instead, can speed up queries significantly.
To have a net benefit however, the database must be suitably configured
for the specific type of data and queries that are most likely expected.
Moreover, different database solutions exist, each of them
tailored to specific types of data.

\begin{figure}[t]
\centering\includegraphics[width=8cm]{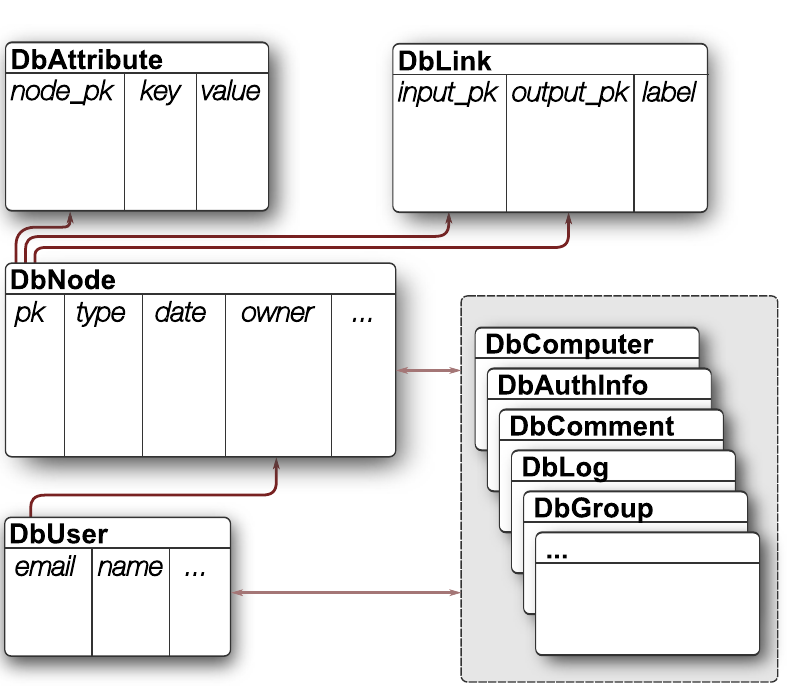}
\caption{\label{fig:schema}
The most relevant tables of the AiiDA database schema. 
The \texttt{DbNode} table contains an entry for each node,
with only a limited set of columns such as the ID (or primary key, PK),
a string identifying the type of node (Calculation, Data, Code,
or a subclass), the creation date, and
the owner (a foreign key to the \texttt{DbUser} table).
All other attributes are stored in the \texttt{DbAttribute} table, as described in the text.
A third \texttt{DbLink} table stores all the links (each link being identified
by the PK of the input and output endpoints, and by a label). 
Other tables exist in the database (to store computers, 
authorization information, comments, log messages, groups, 
...) typically referencing both to nodes and users
(e.g., the comment of a given user on a given node).}
\end{figure}

In this section, we discuss only the details of the
storage solution implemented in AiiDA, and
we defer to~\ref{sec:attributestable}
a discussion on database types (e.g. SQL vs. NoSQL)
and on the reasons for our implementation choices.
After benchmarking different solutions,
we have chosen to adopt a SQL backend for the AiiDA database. 
In particular, MySQL~\cite{mysql} and PostgreSQL~\cite{postgresql}
are fully supported, together with the
file-based backend SQLite~\cite{sqlite} (even if the latter
is not suited for multiple concurrent accesses, and its usage is limited to testing purposes).
The database is complemented by a file repository, where arbitrary files and
directories can be stored, useful for large amounts of data that
do not require direct querying, and is going to be discussed in details later in Sec.~\ref{sec:dbvsfiles}. 
In our implementation, the three main pieces
of information of the DAG (nodes, links,
and attributes) are stored in three
SQL tables, as shown in
Fig.~\ref{fig:schema}.

The main table is called \texttt{DbNode}, where
each entry represents a node in the database.
Only a few static columns are defined: 
an integer identifier (ID), 
that is also the Primary Key (PK) of 
the table; a universally-unique identifier or UUID, 
a ``type'' string to identify the type of node
(\texttt{Calculation}, \texttt{Data}, \texttt{Code},
or one of their subclasses, see Sec.~\ref{sec:orm}). 
A few more columns exist for ``universal''
information such as a label, the creation
and modification time, and the user who
owns the node (a foreign link to the \texttt{DbUser}
table, storing user details).

A second table, \texttt{DbLink}, keeps track of all
directional links between nodes.
Each entry contains the PKs of
the input and output nodes of the link, and a text field for the link label, that
distinguishes the different inputs to a calculation node (e.g., a crystal structure, a set of parameters,
a list of k-points, etc.). 
For instance, link names used for a Quantum ESPRESSO calculation can be seen in Fig.~\ref{fig:simpleprovenance}.

\begin{figure}[t]
\centering\includegraphics[width=8cm]{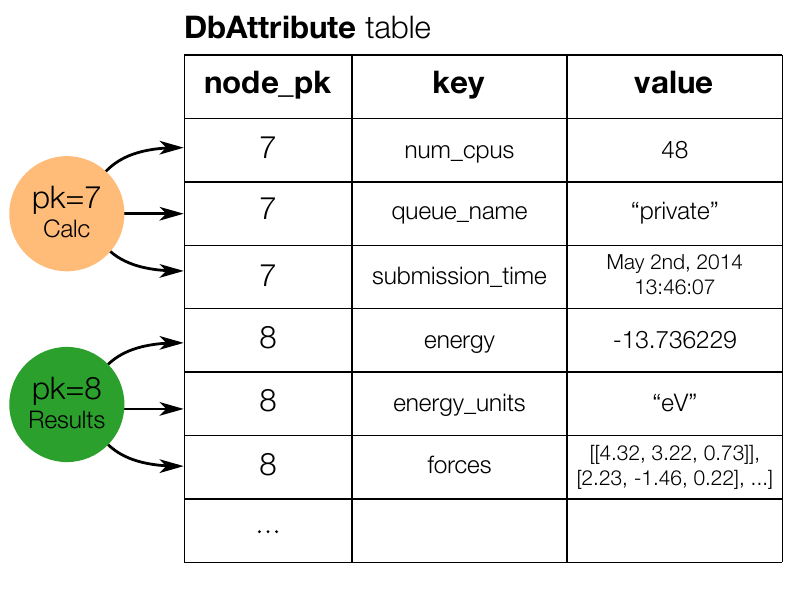}
\caption{\label{fig:nosqlinsql}Arbitrary attribute data in a SQL EAV table
by means of our \texttt{DbAttribute} table. This schema is simplified; the actual
schema implemented in AiiDA is described in \ref{sec:attributestable}.}
\end{figure}

A third table, \texttt{DbAttribute},
is used to store \emph{any} possible attribute that
further characterizes each node.
Some examples of attributes could be: an energy value, a
string for the chemical symbol of each atom in a crystal structure, a
$3\times 3$ matrix for the components of the crystal vectors,
an integer specifying the number of CPUs that we want to use
for a given calculation, \ldots

The \texttt{DbAttribute} table is schematically
represented in Fig.~\ref{fig:nosqlinsql}. 
Each entry represents one attribute of a node, and for 
each attribute we store: the PK
of the node to which this attribute belongs; the \emph{key}, i.e.
a string defining the name of the property that we want to store
(e.g. ``energy'', ``atom\_symbols'', ``lattice\_vectors'',
``num\_cpus'', \ldots); and the \emph{value} of the given property.
Internally, the table has a more complicated schema allowing for 
extended flexibility:
\begin{itemize}
\item Different primitive data types can be stored (booleans, integers,
  real values, strings, dates and times, \ldots)
\item Arbitrary Python dictionaries (sets of \texttt{key=value}
  pairs) and lists can be stored, and any element of the list
  or of the dictionary can be directly and efficiently queried
  (even in case of multiple depth levels: lists of lists,
  lists of dictionaries, \ldots)
\end{itemize}
The technical description of the EAV table
is deferred to~\ref{sec:attributestable}. We emphasize
here that by means of this table we achieve both flexibility, by being able
to store many data types as an attribute, and
preserving query efficiency, since any element in the database can be queried directly
at the database level (making full use of indexes, etc.).

Since each row of the \texttt{DbAttribute} table is an internal
property of a single node, we enforce that attributes 
cannot be modified after the respective 
node has been permanently stored (for example,
we do not want the number of CPUs of a calculation to
be changed after the calculation has been stored and
executed).
However, the user will often find it useful to store 
custom attributes for later search and filtering (e.g., a tag specifying
the type of calculation, the spacegroup of a crystal structure, \ldots).
To this aim, we provide a second table (\texttt{DbExtra})
that is identical to the \texttt{DbAttribute} table
(and therefore it has the same data storage and querying
capabilities). The content of the \texttt{DbExtra} table, though, is not 
used internally by AiiDA, and is at complete disposal of the user.

Besides the three tables \texttt{DbNode}, \texttt{DbLink} and \texttt{DbAttribute} 
that constitute the backbone of the database structure, there are a
few other tables that help data management and organization. The most relevant are:
\begin{itemize}
\item \texttt{DbUser} contains user information (name, email, institution).
\item \texttt{DbGroup} defines groups of nodes to organize and gather
  together calculations belonging to the same
  project, pseudopotentials of the same type, etc.
\item \texttt{DbComputer} stores the list of 
  remote computational resources that can be used to run the simulations.
\item \texttt{DbAuthInfo} stores the authorization information for a given 
  AiiDA user (from the \texttt{DbUser} table) to log in a given computer (from
  the \texttt{DbComputer} table), like the username on the remote computer, etc.
\item \texttt{DbWorkflow}, \texttt{DbWorkflowData}, \texttt{DbWorkflowStep} are
  the tables that store workflow-related information.
\item \texttt{DbPath} is the transitive closure table, described in the next section.
\end{itemize}

\subsection{\label{sec:graphdb}Graph database: querying the provenance}
\begin{figure}[t]
\centering\includegraphics[width=8cm]{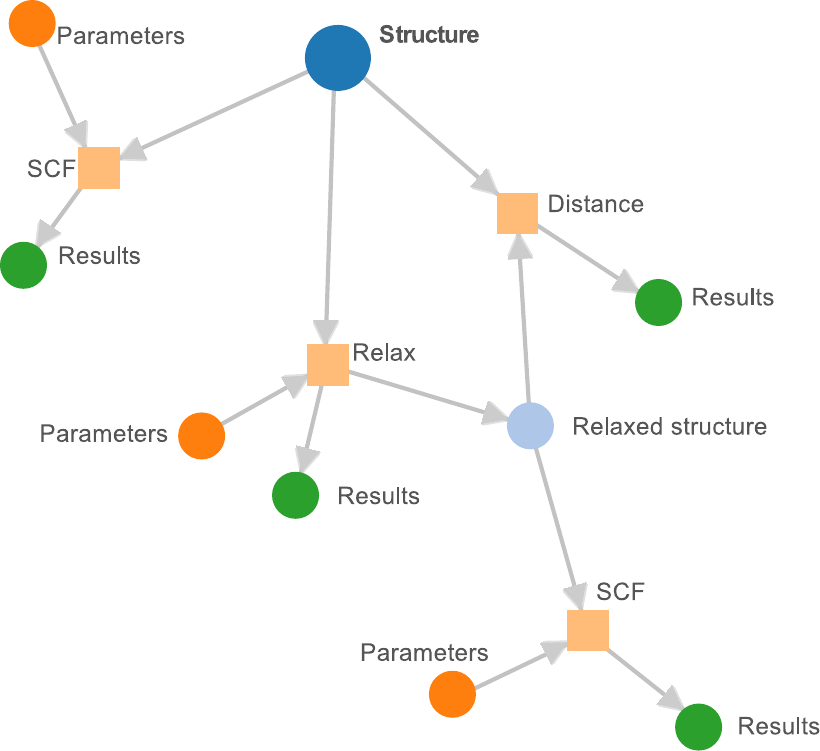}
\caption{\label{fig:graphdbexample}
A schematic example of a possible graph inside the AiiDA database. An initial crystal structure
(blue node on the top of
the figure) is used both for a total energy calculation (SCF) and for a
structural relaxation calculation (Relax). Each calculation has a
set of input parameters (orange dots) and of output results (dark
green dots). The relaxation calculation also produces
another crystal structure, used as input for a new SCF
calculation. Moreover, the two structures in the graph are both
taken as input from a calculation that computes a
suitably-defined ``distance'' between them.}
\end{figure}

An example of a simple graph that can be stored in AiiDA is shown in
Fig.~\ref{fig:graphdbexample}, where four different calculations have been run
with different inputs producing a set of output results, and where some output nodes 
have been used as input of new calculations.
Using the database data model described in the previous section, we can
store DAGs with arbitrary queryable attributes associated to each node.
However, there is another type of query specific to
graph databases, related to the graph connectivity:
given two nodes, to determine the existence of a path connecting them. 
This is particularly relevant for simulations in Materials Science:
typical queries involve searching for crystal structures
with specific computed properties, but the number of intermediate steps
(i.e., \texttt{Calculation} nodes) between a structure data node and
the final result can be large and not even predictable (e.g.,
if multiple restarts are required, \ldots). This type of searches requires in practice
to query the provenance of the data in the database.

Sophisticated and efficient \emph{graph traversal} techniques
have been developed to discover the existence of a path
between two nodes, and graph databases (e.g.,
Neo4j~\cite{neo4j}) implement these functions at the database level, however they
require the use of custom querying languages. 
Instead, we address the graph traversal problem within a SQL database
by incrementally evaluating
a \textbf{transitive closure} table (that we called \texttt{DbPath}).
This table lists the ``paths'', i.e.,
all pair of nodes
(``parent'' and ``child'') that are connected together 
in the graph through a chain of links.
The table is automatically updated every time a link is added,
updated or removed from the \texttt{DbLink} table, by means of database triggers that we have
developed for the three supported backends (SQLite, MySQL and PostgreSQL).
The algorithm for the update of the transitive closure table
has been inspired by Ref.~\cite{Dong:1999} and is described in detail in~\ref{sec:transitiveclosure}.

Obviously, the \texttt{DbPath} table allows for fast queries on the data ``history'',
at the expense of occupying additional disk space for its storage.
The size of the table can in general become very large; however, in Material Science applications, we typically do not have a single dense graph where all nodes are interconnected;
instead, one often creates many small graphs of the type of Fig.~\ref{fig:graphdbexample}. This means that the size of the \texttt{DbPath} table will remain roughly linear in the
number of nodes in the graph.
After benchmarking, we have chosen the solution described above as a good compromise between storage and query efficiency.

\subsection{\label{sec:dbvsfiles}Database vs.\@ file repository}
The storage of attributes in the database discussed previously
allows for query efficiency, at the price
of disk space (for additional information
like indexes, data types, ...) and of efficiency in regular
operations (retrieving a large list from the database
is slower than retrieving a file containing the same list).
For large matrices, therefore, a threshold exists above
which the advantages of faster query speed are
not justified anymore. Moreover, the single
entries of a large matrix (like the electron charge density
discretised on a grid) are often of little interest
for direct querying.
In such cases it is convenient to store data in a file and rely on the file system for I/O access. 
This is especially appropriate when the data should be used as an input to another calculation and a fast read access is required.

For these reasons, AiiDA complements the database storage with a
file repository (see Fig.~\ref{fig:infrastructure}), confined 
within a folder configured during the setup phase of AiiDA.
Every time a new node is created, AiiDA automatically
creates a subfolder for the specific node.
Any file, if present, associated with the node will be stored therein.

Storing information as files or in 
the database is left as a choice for the developers of 
the specific subclass of the \texttt{Data}
of \texttt{Calculation} nodes (plugins are discussed in Sec.~\ref{sec:orm}),
but in order to maximize efficiency, it should follow the guideline discussed above.
Some examples of the choices made for some AiiDA \texttt{Data} plugins can be
found in Sec.~\ref{sec:plugins}.

\section{\label{sec:environment}The scientific environment in AiiDA}
\subsection{\label{sec:orm}The ORM}
AiiDA is written in Python, a powerful object-oriented 
language. The materials simulation community has been moving
towards Python in recent years, due to its simplicity 
and the large availability of libraries for visualization, text parsing, and
scientific data processing~\cite{Jain:2011,Bahn:2002,spglib}.
An important component of the AiiDA API is the Object-Relational Mapper (ORM),
which exposes to the user only a very intuitive Python interface
to manage the database.
The main class of the AiiDA ORM, \texttt{Node}, is used to represent any
node in the graph. Each instance of the \texttt{Node} class internally
uses Django to perform database operations, and complements it with
methods for accessing the file repository. The low-level interaction
with the database uses the Django framework~\cite{django}.

The main functionalities of the \texttt{Node} class are:
\begin{itemize}
\item It provides a direct access to the node attributes, accessible
  as a Python dictionary by using the
  \texttt{node.attrs()} method.
  The method also properly recreates
  lists and dictionaries that were stored in expanded format at
  the database level (lists and dictionaries are not natively supported in the chosen databases), as described in~\ref{sec:attributestable}.
  Similar methods allow the user to read and write user-defined
  attributes in the \texttt{DbExtra} table.
\item It provides direct access to the repository folder
  containing the files associated to each \texttt{Node}. \item It provides a caching mechanism that allows the user to create
  and use the \texttt{Node} even before storing it in the database or
  on the repository folder, by keeping the files in a temporary
  sandbox folder, and the attributes in memory. 
  This is particularly useful to test the generation of the input
  files by AiiDA without the need to store test data in the database.
  Only after the
  \texttt{node.store()} call, all the data is
  permanently stored in the AiiDA database and repository and no
  further modifications are allowed.
  A similar caching mechanism has also been implemented to keep track of
  links between nodes before storing.
\item It provides an interface for querying nodes with
  specific attributes or with specific attribute values, or nodes with given inputs or outputs, etc.
\item It provides methods (\texttt{.inp} and \texttt{.out}) to get the list of inputs and
outputs of a node and similarly the list of all
parent and child nodes using the transitive closure table. 
\end{itemize}

\subsection{\label{sec:plugins}The plugin interface}
To support a new type of calculation
or a new kind of data (e.g. a band structure, a
charge density, a set of files, a list of parameters, \ldots),
one simply needs to write an AiiDA plugin. A plugin
is simply a python module file, containing the definition
of a subclass of the AiiDA classes, sitting in an
appropriate folder; AiiDA automatically detects the
new module and uses it.

All different types of nodes are implemented as 
subclasses of the \texttt{Node} class. At a first
subclass level we have the three main 
node types: \texttt{Calculation}, \texttt{Code} and \texttt{Data}.
Each of them is further subclassed by plugins to provide specific functionalities.
In particular, instances of \texttt{Code}
represent a specific executable file
installed on a given machine (in the current
implementation, there are no further subclasses).
Each subclass of \texttt{Calculation}, instead, supports a
new simulation software and contains the code needed
to generate the software-specific
input files starting from the information stored
in the AiiDA database. Moreover, it can
also provide a set of software-dependent methods (like
\texttt{calc.restart()}, \ldots)
that make it easier for the user to perform 
routine operations.
Finally, the \texttt{Data} class has a subclass
for each different type of data that the user wants to represent.
The specific subclass implementation determines
the possible user operations on the data, and whether the information
is going to be stored in the database as attributes or in the file repository.
We report here a description of some of the most relevant \texttt{Data}
subclasses distributed with AiiDA:
\begin{itemize}
\item \texttt{ArrayData}: it is used to store (large) arrays. Each 
  array is stored on disk as a binary, portable, compressed
  file using the Python \texttt{numpy} module~\cite{vanderWalt:2011}.
  Some attributes are stored in the \texttt{DbAttribute} table for
  fast querying (like the array name and its size). Subclasses
  use the same storage model, but define specific methods to 
  create and read the data (e.g., the \texttt{KpointsData} class 
  has methods to detect a structure cell,
  build the list of special $k$-points in $k$ space
  and create paths of $k$-points, suitable for plotting band
  structures, using for instance the standard paths listed in Ref.~\cite{Setyawan:2010}).
\item \texttt{ParameterData}: it is used to store the content of a 
  Python dictionary in the database.
  Each key/value pair is stored as an attribute
  in the \texttt{DbAttribute} table, and no files are stored in the repository.
\item \texttt{RemoteData}: this node represents a ``link'' to a directory
  on a remote computer. It is used for instance
  to save a reference to the scratch folder on the remote computer in which the
  calculation was run, and acts as a placeholder 
  in the database to keep the full data provenance, for instance
  if a calculation is restarted using the content
  of that remote folder.
  No files are written in the AiiDA repository,
  but the remote directory absolute path is stored as an attribute.
\item \texttt{FolderData}: this node represents a folder with files. At
  variance with \texttt{RemoteData}, files are
  stored permanently in the AiiDA repository (e.g.,
  the outputs of a finished calculation retrieved from the
  remote computer).
\item \texttt{StructureData}: this node represents a crystal structure.
  The $3\times 3$ coordinates of the lattice vectors,
  the list of atoms and their coordinates, and any other information
  (atomic masses, \ldots) are saved as attributes for easy querying.
  (For very large structures, a different
  data model may be more efficient.)
  Methods are provided for standard operations like getting
  the list of atoms, setting their positions and masses,
  converting structures to and from other formats (e.g.\@ the
  \texttt{Atoms} class of the ASE Atomistic
  Simulation Environment~\cite{Bahn:2002}), obtaining the structure from an external database (like ICSD~\cite{icsd} or COD~\cite{grazulis:2012}), getting the spacegroup
  using SPGlib~\cite{spglib}, etc. 
\end{itemize}

Finally, we emphasize that the plugin interface is not limited
to the ORM, and a similar plugin-based
approach applies to other AiiDA components, like the
connection transport channel and the schedulers (as discussed in
Sec.~\ref{sec:transportandscheduler}).

\subsection{\label{sec:verdi}User interaction with AiiDA}
We provide a few different interfaces to interact with AiiDA.
The most commonly used is the \texttt{verdi} command line utility.
This executable exposes on the
command line a set of very common operations, such as performing the
first installation; reconfiguring AiiDA; listing or creating codes,
computers and calculations; killing a calculation; starting/stopping the
daemon, \ldots{}
The \texttt{verdi} tool is complemented by a Bash completion feature to 
provide suggestions on valid commands by pressing the TAB key.
Moreover, an inline help provides a list of existing
commands and a brief description for each of them.
The advantage of \texttt{verdi} is to expose
basic operations to the user without requiring any knowledge of Python
or other languages.

In order to access the full AiiDA API, however, the best approach
is to write Python scripts. The only difference with respect to 
standard python scripts is that a
special function \texttt{aiida.load\_dbenv()} 
needs to be called at the beginning of the file
to instructs Python to
properly load the database.
Once this call has been made, any class from the
\texttt{aiida} package can be loaded and used.
If the users do not want do explicitly call the
\texttt{aiida.load\_dbenv()} call in the python code,
then they can run the script using the \texttt{verdi run} command.
In this case, the AiiDA environment and some default AiiDA classes
are automatically loaded before executing the script.

A third interface is the interactive python shell that can be loaded
using the command \texttt{verdi shell}. The shell is based
on IPython~\cite{Perez:2007} and has the advantage to
automatically load the database environment; at the same time, it
already imports by default some of the most useful classes (e.g.
\texttt{Node}, \texttt{Calculation}, \texttt{Data}, \texttt{Group}, 
\texttt{Computer}, \ldots) so that they are directly
available to the user. TAB completion is available and very useful to
discover methods and attributes. Moreover,
the documentation of each class or
method (written as Python ``docstrings'') is directly accessible.

\subsection{\label{sec:scientificworkflows}Scientific workflows}
As introduced in Sec.~\ref{sec:requirements},
many tasks in scientific research are standard and frequently repeated,
and typically require multiple steps to be run in sequence.  
Common use cases are parameter convergence, restarts in 
molecular dynamics simulations, multi-scale simulations, data-mining analysis, 
and other situations when results of calculations with one code are used as
inputs for different codes. 
In such cases, it is beneficial to have a system that 
encodes the workflow and manages its
execution~\cite{galaxy,kepler,knime,vistrails,Wolstencroft:2013}.

In order to fully integrate the workflows within the ADES model, 
we implement a custom engine into AiiDA, by which the user can 
interact with all AiiDA components via the API.
This engine is generic and can be used to define any computational workflow.
Specific automation schemes, crafted for selected applications 
(equation of states, phonons, etc...) are implemented within each workflow
and can be developed directly by the users.

AiiDA workflows are subdivided into a number of steps.
One or more calculations can be associated to each step;
these are considered to be independent and are launched in parallel.
Instead, different steps are executed sequentially and the execution
order is specified by ``forward'' dependency relationships: 
In other words, each ``parent'' step must specify the step to be executed next.
The execution of the ``child'' step is delayed until
all calculations associated to the parent have completed.
We emphasize that dependency relationships are defined only between steps.
Dependencies between calculations are implicit, with the advantage of allowing for
both parallel and serial simulation streams.

Within a step, beside creating calculations and
associating them to the current step, any other 
Python (and AiiDA) command can be executed. This gives maximum flexibility
to define complex workflow logics, especially if the 
outputs of a calculation in a parent step require some processing
to be converted to the new calculation inputs.
Moreover, a step can define itself as the next
step to be executed, providing support for 
loops (even conditional ones, where the number of iterations
depends on the calculations results).

A key feature, modularity, completes AiiDA workflows: 
within each step, the user can associate 
not only calculations, but also subworkflows.
The advantage is the possibility to reuse existing workflows
that perform specific tasks, so as to develop only missing features. 
For instance, let us assume that we developed a workflow ``A'' that
performs a DFT calculation implementing code-specific
restart and recover routines in order to make sure convergence is achieved.
Then, a workflow ``B'' that calculates the energy of a
crystal at different volumes (to obtain its equation of state)
does not need to reimplement the same logic, but
will just reuse ``A'' as a subworkflow.
``B'' can in turn become a subworkflow of a higher-level workflow
``C'' that, for instance, compares the equation of state calculated
at different levels of approximation.
The combination of parallel and serial execution, conditional loops,
and modularity, makes the workflows general enough to support
any algorithm. 

From the implementation point of view,
the AiiDA workflow engine is provided by a
generic \texttt{Workflow} class, that can be inherited
to define a specific workflow implementation.
Workflow steps are special class methods
identified by the \texttt{@step} decorator.
The base class also provides dedicated methods to associate 
calculations and workflows to the current step.
In every step, a call to the \texttt{self.next()}
method is used to define dependencies between steps. This method
accepts as a parameter the name of the following step. The
name is stored in the database, and the corresponding
method is executed by the daemon only when all calculations
and subworkflows of the current step have finished.

In analogy with calculations, AiiDA uses states to
keep track of workflows and workflow steps 
(\texttt{RUNNING}, \texttt{FINISHED}, \ldots).
The AiiDA daemon handles all the workflow operations
(submission of each calculation, 
step advancement, script loading, error reporting,
\ldots) and the transitions between different workflow states.

In the long term, we envision integrating into AiiDA 
many new and existing methods in the form of workflows
(e.g., training interatomic potentials~\cite{Bartok:2010}, crystal structure
prediction algorithms~\cite{Glass:2006,Pickard:2011}, \ldots),
so that the researcher can focus on materials science
and delegate to AiiDA the management of
remote computers, the appropriate choice of code-specific parameters,
and dealing with code-specific errors or restarts.

\subsection{Querying}
A relevant aspect of a high-level scientific environment is the possibility
of running queries on the data stored in AiiDA without the need to know a
specific (and typically complex)
query language. To this aim, we have developed a Python class, called the \texttt{QueryTool},
to specify in a high-level format the query to run. For instance, it is possible to
specify a filter on the node type
(e.g., to get only crystal structures); to filter by the value of a specific attribute or \texttt{DbExtra} table entry (e.g., to select
structures with a specific spacegroup); or to filter by a specific attribute in 
one of the linked nodes (e.g., to get structures on which a total-energy calculation was run,
and the energy was lower than a given threshold). Queries can also take advantage of the
transitive closure table, setting filters on attributes of nodes connected to the one of 
interest by an unknown number of intermediate links (e.g., if the result of the
calculation was obtained after an unknown number of restarts).
As an example, a complex query that can be
run with AiiDA is: \emph{give me all crystal structures containing iron and oxygen
and with a cell volume larger than $X$~\AA$^3$,
that I used as input for a sequence of calculations with codes $Y$ and $Z$
to obtain the phonon band structure, for which the lowest phonon frequency that was obtained is positive, and for which the DFT exchange-correlation functional used was LDA.} Other complex queries can be specified using the \texttt{QueryTool} in a format that is easy both to write and to read. 

\subsection{\label{sec:docs}Documentation and unit testing}
\begin{figure}[t]
\centering\includegraphics[width=8cm]{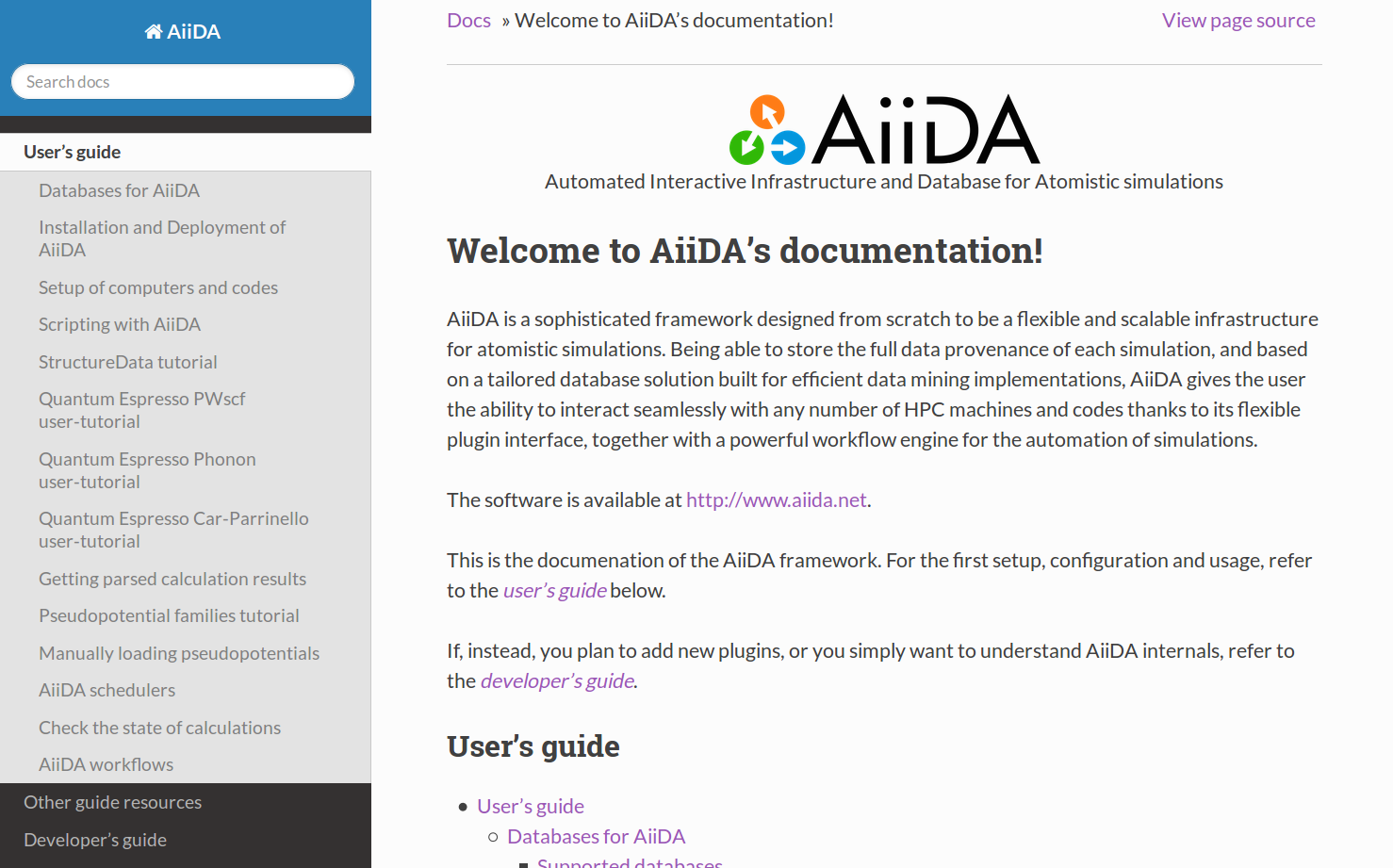}
\caption{\label{fig:docs}The first page of the AiiDA documentation. Different sections
are aimed at the end users (with examples and tutorials) and at developers (with 
the documentation of classes and methods, and tutorials for new plugin development).}
\end{figure}

A key component of a productive scientific environment is a complete
and accurate code documentation. For this reason, AiiDA is complemented
by an extensive documentation.
Each class, method and function has a
Python docstring that describes what the function does, the arguments,
the return values and the exceptions raised. 
These docstrings are accessible via the interactive shell, but are
also compiled using the Sphinx~\cite{sphinx} documentation engine into
a comprehensive set of HTML pages (see Fig.~\ref{fig:docs}).
These pages are distributed with the code (in the \texttt{docs/} 
subfolder) and also available online 
at \url{http://aiida-core.readthedocs.org}.
Moreover, we also provide in the same format,
using Sphinx, a complete user documentation of the different
functionalities, supported databases, classes and codes,
together with tutorials covering the installation
phase, the launch of calculations and workflows, 
data analysis,~\ldots{} The user guide
is also complemented by a developer's guide that
documents the API and contains tutorials for the development of new plugins.

Moreover, to simplify the code maintenance, we implement a
comprehensive set of unit tests (using the Python
\texttt{unittest} module), covering the different components
of AiiDA described in Fig.~\ref{fig:infrastructure}. 

\section{\label{sec:sharing}Sharing in AiiDA}
The fourth pillar introduced in Sec.~\ref{sec:requirements} aims at enabling a social ecosystem where it becomes easy to
share tools and results, such as data, codes, and workflows.
In order to preserve the authorship and privacy of the data of each researcher,
we implemented a model in which each user or group of users
can install their own local AiiDA instance. All data and calculations
are accessible only to people with direct access to the instance and therefore remain private. In order to enable sharing of
the database (or parts of it) with collaborators, 
we provide functionality to export a portion
of the database to a file, and then to import it in a
different instance of AiiDA. In this way several groups in a collaboration may contribute to a common repository, open to the entire project, while retaining their private portions as needed. This approach simplifies issues of user- and group-level security. To avoid conflicts during this procedure, AiiDA
assigns a Universally Unique IDentifier (UUID) to each node 
as soon as it is locally created.
In fact, while for internal database usage an auto-incrementing integer PK is the most
efficient solution to refer to a database row,
the PK is not preserved when a node is transferred to a different database. Instead, the first node to be created will always have PK=1, the second PK=2, and so on.
A UUID is instead a hexadecimal string that may look like the following:
\texttt{e3d21365-7c55-4658-a5ae-6122b40ad04d}.
The specifications for creating a UUID are defined in
RFC~4122 by IETF~\cite{rfc4122}, and they guarantee that 
the probability that two different UUIDs generated in different space and
time locations can be assumed to be zero.
In this way, we can use the UUID to identify the existence of a node in the DB; 
at import time, it is only verified whether a node with the same UUID
already exists before including its contents in the database.
We envision one or several centralized repositories, available to the public,
that collect results from different groups. Researchers
will be able to share results
with colleagues by exchanging just the UUID of the nodes
stored on the public repository, rather than sending
actual scripts or files. In this collaborative environment
the adoption of new codes, the comparison of results,
and data disclosure and reproducibility become straightforward, realizing the ``social ecosystem'' discussed in Sec.~\ref{sec:requirements} and facilitating the reusability of data, codes and workflows. We emphasize that only after having a locally deployed automation infrastructure like AiiDA it becomes feasible
to populate public repositories efficiently and in a uniform format, because the user effort to uniform the data, prepare it and upload it is reduced to the minimum.

The standardization of the formats produced by
different codes is necessary to
maximize the effectiveness of data sharing. This task is outside
the scope of the AiiDA project, but we encourage existing and future
standardization efforts by the developers of the simulation codes.
At the same time, within AiiDA we provide a set of ``default'' 
plugins for the most common data structures (crystal structures, paths of $k-$points
in reciprocal space, \ldots), which can be seamlessly reused in different codes.
Implemented classes provide importers and exporters
from/to common file formats, so that it is possible to
exchange the data also with other infrastructures and
repositories that use different data formats.
We also emphasize that it is possible to write workflows that perform high-level tasks common to a variety of codes, such as structure optimization or molecular dynamics, even before a standardization of data formats takes place.

Moreover, to encourage the development of the repository pipelines
discussed in the ``Sharing'' pillar, we 
define in AiiDA a common API for importers and exporters of crystal structures
(in the \texttt{aiida.tools} package). Also in this case, repositories can be
supported by plugin subclasses; plugins for importing structures from 
some common databases, such as ICSD~\cite{icsd}, COD~\cite{grazulis:2012},
MPOD~\cite{mpod} and TCOD~\cite{tcod} are already distributed with the code.

\section{Codes and data types supported out of the box}
In the first stage of the development of AiiDA,
we mainly focused on the infrastructure and its components (scheduler
management, database, \ldots) rather than writing a large number of
plugins.

However, we already provide with AiiDA 
fully-functional plugins for most
codes of the Quantum ESPRESSO package~\cite{Giannozzi:2009}, namely
\texttt{pw.x}, \texttt{ph.x}, \texttt{cp.x}, and many of the post-processing
tools such as \texttt{matdyn.x}, \texttt{q2r.x}, \ldots 
These plugins expose the full functionality of the respective codes,
and moreover include not only the input generation routines, but also
output parsers.
The type and format of the input nodes required by Quantum ESPRESSO is
shown in Fig.~\ref{fig:simpleprovenance} and described in 
detail in the code documentation. 
We also provide a \texttt{PwImmigrant} class to import 
\texttt{pw.x} calculations that were already run before 
the adoption of AiiDA.

Similarly to calculations, the user can support new custom tailored data types by
defining a new data format. However, to facilitate the standardization
of the most common data types, we provide with AiiDA a set
of \texttt{Data} subclasses. Some examples (already described in
Sec.~\ref{sec:plugins}) include the \texttt{StructureData} for crystal
structures (supporting also non-periodic systems, vacancies and
alloys), the \texttt{KpointsData} for paths in $k-$space, the \texttt{BandsData} for
band-structure data, the \texttt{ArrayData} for generic arrays.
Moreover, we also provide a specific class to support pseudopotentials
in the UPF format, that is the format used by Quantum ESPRESSO. 
Specific routines allow the user to upload a single pseudopotential,
or a family of pseudopotentials into the database. At upload time, the
pseudopotentials are parsed to discover to which element they
correspond, and an attribute is added for later querying. Moreover,
the MD5 checksum of the file is calculated and stored, to avoid to
store twice the same pseudopotential in the database.
The \texttt{verdi data upf uploadfamily} command can be used to take all 
UPF files contained in a folder, possibly of the same type (e.g. ultrasoft with
PBE functional, \ldots) and group them in a ``family''.
In this way, when running a Quantum ESPRESSO
simulation, one can ask AiiDA to simply select the appropriate
pseudopotentials from a family, rather than manually selecting the
pseudopotentials one by one.

We include support also for other codes,
in particular all the \emph{codtools}~\cite{Grazulis:2015},
a set of tools for processing, filtering and correcting CIF files (the 
\emph{de-facto} standard format for crystal structures).
Furthermore, a basic support to the GPAW code~\cite{Mortensen:2005} is provided and,
since GPAW runs via the ASE interface~\cite{Bahn:2002}, the same plugin can
be used to access a large family of simulation codes already supported
by ASE, like VASP, ABINIT, Gaussian, \ldots The full list can be found
on the ASE webpage.

\section{\label{sec:examples}Examples of applications}
In this section, we summarize some examples in the field of materials science
in which AiiDA has already been used as the platform for
managing simulations.

\subsection{\label{sec:exverification}Verification and validation}
Tests for verification and validation purposes can often be automated.
One typical example in the field is the test of pseudopotentials for accuracy and transferability. AiiDA workflows can be developed to perform standardized tests with minimal user intervention, as soon as a new pseudopotential file (or family
of pseudopotentials) is available. 
If a pseudopotential is discovered to be inaccurate,
one can easily find and flag all calculations that were performed with it. One can then easily
run again all affected calculations with an improved pseudopotential
with minimal effort, since each calculation stored in AiiDA is 
reproducible and the provenance of the data is available.

As an example, we implemented a workflow (using Quantum ESPRESSO as the
simulation engine) to perform the Crystalline Monoatomic Solid Test (CMST) 
with the 15-point protocol described in Ref.~\cite{Kucukbenli:2014}.
 The workflow, starting from an initial guess of the lattice parameter
$a_0$, iteratively refines the value of $a_0$ by fitting a Birch--Murnaghan
equation of state to 15 calculations performed at different volumes.
An automatic restart procedure is implemented as a subworkflow
to automatically relaunch any calculation that crashed 
using different numerical parameters, as discussed in
Sec.~\ref{sec:scientificworkflows}
(e.g. using a different initial guess for the wavefunctions
or a different convergence algorithm).
We have been able to reproduce the results
for the PSlibrary~\cite{pslibrary} family of pseudopotentials
reported in Ref.~\cite{Kucukbenli:2014},
and we computed the same quantities also for the GBRV
pseudopotentials~\cite{Garrity:2014} using the suggested cutoffs (40~Ry
for the wavefunctions and 200~Ry for the charge density), a shifted
$32\times32\times32$ $k-$mesh and a Marzari-Vanderbilt smearing~\cite{Marzari:1999} of 2~mRy.
The results are reported in Table~\ref{tab:cmst}.

\begin{table}[t]
\centering\footnotesize
\begin{tabular}{lrrp{0.2cm}rrrp{0.2cm}rrr}
\hline & \multicolumn{2}{c}{FCC} && \multicolumn{3}{c}{BCC} && \multicolumn{3}{c}{SC}\\
Element & $a_0$ & $B_0$ && $a_0$ & $B_0$ & $\Delta E^{fcc}$ && $a_0$ & $B_0$ & $\Delta E^{fcc}$\\\hline
Ag & $  4.151$ & $ 90.7$ && $  3.302$ & $ 88.7$ & $  0.031$ && $  2.752$ & $ 66.4$ & $  0.332$\\
Ba & $  6.357$ & $  8.1$ && $  5.024$ & $  8.7$ & $ -0.017$ && $  3.955$ & $  7.8$ & $  0.287$\\
Be & $  3.163$ & $119.8$ && $  2.506$ & $125.2$ & $  0.018$ && $  2.176$ & $ 76.5$ & $  0.920$\\
Ca & $  5.527$ & $ 17.4$ && $  4.384$ & $ 15.5$ & $  0.017$ && $  3.515$ & $ 10.7$ & $  0.393$\\
Cd & $  4.509$ & $ 41.9$ && $  3.609$ & $ 36.4$ & $  0.051$ && $  3.000$ & $ 30.4$ & $  0.114$\\
Co & $  3.452$ & $257.7$ && $  2.760$ & $242.8$ & $  0.255$ && $  2.281$ & $186.6$ & $  0.864$\\
Cr & $  3.625$ & $236.6$ && $  2.850$ & $257.4$ & $ -0.393$ && $  2.342$ & $186.2$ & $  0.632$\\
Cs & $  7.770$ & $  2.0$ && $  6.161$ & $  2.0$ & $  0.001$ && $  5.043$ & $  1.6$ & $  0.093$\\
Cu & $  3.633$ & $141.7$ && $  2.888$ & $139.7$ & $  0.035$ && $  2.409$ & $104.1$ & $  0.469$\\
Fe & $  3.448$ & $287.0$ && $  2.757$ & $270.4$ & $  0.317$ && $  2.266$ & $209.4$ & $  0.961$\\
K & $  6.664$ & $  3.6$ && $  5.282$ & $  3.6$ & $  0.000$ && $  4.295$ & $  2.9$ & $  0.105$\\
Li & $  4.325$ & $ 13.8$ && $  3.435$ & $ 13.9$ & $  0.002$ && $  2.732$ & $ 12.2$ & $  0.121$\\
Mg & $  4.524$ & $ 35.2$ && $  3.580$ & $ 35.4$ & $  0.016$ && $  3.022$ & $ 22.7$ & $  0.367$\\
Mn & $  3.505$ & $279.7$ && $  2.785$ & $278.1$ & $  0.079$ && $  2.284$ & $211.9$ & $  0.862$\\
Mo & $  4.000$ & $238.0$ && $  3.160$ & $258.8$ & $ -0.426$ && $  2.599$ & $187.6$ & $  0.733$\\
Na & $  5.293$ & $  7.7$ && $  4.198$ & $  7.7$ & $  0.000$ && $  3.412$ & $  6.1$ & $  0.119$\\
Nb & $  4.216$ & $164.1$ && $  3.309$ & $172.2$ & $ -0.324$ && $  2.719$ & $130.0$ & $  0.664$\\
Ni & $  3.510$ & $204.5$ && $  2.792$ & $200.4$ & $  0.054$ && $  2.323$ & $148.3$ & $  0.668$\\
Pd & $  3.943$ & $170.2$ && $  3.137$ & $167.3$ & $  0.044$ && $  2.615$ & $123.8$ & $  0.516$\\
Rb & $  7.147$ & $  2.8$ && $  5.667$ & $  2.8$ & $  0.001$ && $  4.622$ & $  2.2$ & $  0.095$\\
Rh & $  3.832$ & $256.3$ && $  3.071$ & $235.5$ & $  0.357$ && $  2.538$ & $187.5$ & $  0.798$\\
Ru & $  3.809$ & $307.2$ && $  3.051$ & $281.2$ & $  0.514$ && $  2.510$ & $221.3$ & $  1.025$\\
Sc & $  4.618$ & $ 51.3$ && $  3.676$ & $ 53.4$ & $  0.056$ && $  2.967$ & $ 35.2$ & $  0.721$\\
Sr & $  6.020$ & $ 11.3$ && $  4.756$ & $ 12.1$ & $  0.005$ && $  3.853$ & $  7.0$ & $  0.385$\\
Tc & $  3.870$ & $298.4$ && $  3.080$ & $293.1$ & $  0.183$ && $  2.531$ & $223.6$ & $  0.977$\\
Ti & $  4.113$ & $107.6$ && $  3.255$ & $106.0$ & $  0.053$ && $  2.640$ & $ 76.7$ & $  0.780$\\
V & $  3.816$ & $176.5$ && $  2.996$ & $183.5$ & $ -0.244$ && $  2.448$ & $136.8$ & $  0.607$\\
Y & $  5.060$ & $ 39.4$ && $  4.039$ & $ 39.0$ & $  0.098$ && $  3.262$ & $ 25.8$ & $  0.773$\\
Zn & $  3.931$ & $ 69.2$ && $  3.133$ & $ 63.4$ & $  0.063$ && $  2.630$ & $ 48.3$ & $  0.203$\\
Zr & $  4.526$ & $ 90.7$ && $  3.573$ & $ 88.0$ & $  0.045$ && $  2.913$ & $ 68.9$ & $  0.839$\\\hline
\end{tabular}
\caption{\label{tab:cmst}Results of the fully automated CMST for the PBE GBRV v.1.2
pseudopotentials~\cite{Garrity:2014} on 30 elemental solids in different crystal structures,
using the 15-point protocol of Ref.~\cite{Kucukbenli:2014}, as implemented in our AiiDA workflow. $\Delta E^{fcc}$ indicates the total energy difference between the crystal structure under consideration (BCC or SC) and the FCC structure. Lattice parameters $a_0$ are in \AA{}, bulk moduli $B_0$ in GPa, energy differences in eV.}
\end{table}

\subsection{\label{sec:exfunctional}New functional development}
When developing new methods or functionals, one often needs to
benchmark multiple versions of a simulation code.
It is crucial to record the code performance and the data accuracy, besides the
information needed to reproduce results, in order to detect
code errors and bypass bottlenecks. Moreover, parametric
sweep tests are often performed on a set of training structures to collect
statistics for uncertainty quantification analysis and to
assess transferability of new methods.
As a practical example, AiiDA was used for the development and verification
of the Koopmans-compliant functionals by Borghi \emph{et al.}
in Ref.~\cite{Borghi:2014}.

\subsection{\label{sec:experovksites}High-throughput material screening}
Automated workflows are commonly used for high-throughput 
materials screening, for instance to explore
the chemical composition space of a given crystal structure by changing
the elements that form the structure.
As an example, we are investigating
ABO$_3$ perovskites using automated workflows to calculate finite-temperature
properties in the quasi-harmonic approximation, where intermediate steps
include calculations of equations of state and phonon spectra
at different volumes for every material.

\subsection{\label{sec:exdatabases}Open repositories}
Availability of open repositories is necessary for speeding
up scientific discoveries and the development of new materials. 
The task of uploading calculations to a repository and their reproducibility are made simple by AiiDA, because provenance is captured
in the data model. Therefore, it is straightforward to create new cloud-based repositories of results for public access,
or to upload the results to existing 
open databases, like the TCOD~\cite{tcod}, for which
we develop automated exporters for the computed structures as well as the full tree of calculations that generated them.

\section{\label{sec:conclusions}Conclusions}
In this paper we postulate and discuss the fundamental features required by
an infrastructure for computational science, summarized
in the four pillars of Automation, Data, Environment and
Sharing. We then present the software platform AiiDA (http://www.aiida.net), discussing
in detail its implementation and how these requirements are addressed. 
The core of the platform is represented by the AiiDA API,
an abstract layer of 
Python custom inheritable classes that exposes the user in an intuitive manner to the main AiiDA
\texttt{Calculation}, \texttt{Data} and \texttt{Code} objects.
Using the ORM, objects can be created, modified and queried
via a high-level interface which is agnostic of the detailed storage solution or of the SQL query language.
Data, calculations and their results are safeguarded in a tailored storage formed both by repository folders and an SQL database (PostgreSQL, MySQL and SQLite are natively supported) with a schema that supports storage 
and query of directed acyclic graphs and general attributes (numbers, strings, lists, dictionaries).
Heterogeneous data can be accommodated thanks to the use of entity-attribute-value (EAV) tables, and
graph traversal is optimized by automatically updating a transitive-closure table that can be directly queried to assess node connectivity. The
tight coupling of automation and storage enables data flows to be tracked and safeguarded in such a tailored solution. 
A flexible workflow engine is made available for the definition and execution of
complex sequences of calculations, managed by a daemon, without the need of
direct user intervention. Implementation of multiple connection protocols let the software interact with remote computational resources and submit calculations to different job schedulers.
A modular plugin design allows for seamless integration of additional connection protocols and schedulers, or 
the inclusion or extension to different codes and computational solutions. We believe that such platform will be key to enabling researchers to accelerate their process in computational sciences, removing many of the error-prone details and technicalities of the simulations, while supporting validation, reproducibility, and ultimately an open-access model to computational efforts.

\section*{Acknowledgments}
The authors gratefully acknowledge help from multiple collaborators:
Nikolai Zarkevich, Katharine Fang, Kenneth Schumacher, Wen Huang, Alexei Ossikine and Gopal Narayan for helping implement versions of an earlier infrastructure design; Christoph Koch for many useful discussions and for suggesting the adoption of the transitive closure table; Nicolas Mounet for developing Quantum ESPRESSO workflows and Andrius Merkys for developing import and export functionality for repositories, in particular COD and TCOD, and both for significant improvements to the infrastructure; Eric Hontz for developing import functionalities for existing Quantum ESPRESSO simulations and workflows; Philippe Schwaller for the ICSD importer; Valentin Bersier for the Web interface; Marco Dorigo for the implementation of the SGE scheduler plugin; Giovanni Borghi, Ivano Castelli, Marco Gibertini, Leonid Kahle and Prateek Mehta for acting as early beta-testers and for providing user feedback. 

This work has been supported, in chronological order, by Robert Bosch LLC; by the Laboratory of Theory and Simulation of Materials (THEOS); by the Swiss National Centre for Competence in Research on ``Computational Design and Discovery of Novel Materials'' (NCCR MARVEL); by computing resources at the Oak Ridge Leadership Computing Facility at the Oak Ridge National Laboratory, which is supported by the Office of Science of the U.S. Department of Energy under Contract No. DE-AC05-00OR22725, within the INCITE project with ID mat045; and by the Swiss National Supercomputing Centre (CSCS) under project IDs THEOS s337, MARVEL mr1 and CHRONOS ch3.
\appendix

\section{\label{sec:ourbackend}The AiiDA storage implementation -- technical details}

\subsection{\label{sec:attributestable}Attributes in a EAV table}
The most widespread databases are of the relational kind, based on the
Structured Query Language (SQL). 
In a SQL database, the information
is stored in a set of tables, where each entry is a row of the table,
and each column is an entry property. 
The number and type of columns is predefined in
the \emph{database schema} when the database is created, and is not
changed during the database usage.
This requirement can be a strong limitation for applications
that require a variable set of properties to be stored.
This has led to the development of a number of database 
systems with different underlying models for the data storage; 
all these solutions fall under the generic name of
NoSQL solutions (a few popular examples include Cassandra~\cite{cassandra},
CouchDB~\cite{couchdb}, MongoDB~\cite{mongodb}, Neo4j~\cite{neo4j}, 
but many more exist).

The final choice of the backend depends on a tradeoff between usability,
storage speed and size, and query efficiency.
These aspects strongly depend in turn on the type of
stored data and on the typically run queries.
After comparing different solutions
we decided to use a SQL backend with a suitably designed
schema for the two attributes tables (\texttt{DbAttribute} and
\texttt{DbExtra}). These two are modified 
Entity--Attribute--Value (EAV) tables.
Beside the \texttt{node\_pk} column, pointing
to a node in the \texttt{DbNode} table, and
the \texttt{key} column to store the name of the
attribute, we also have a few columns to store
the attribute value. In order to have efficient queries,
we decided to have one column for each
of the basic datatypes that we want to consider: boolean 
values, integers, reals, strings and date/times.
These five columns are complemented by a \texttt{datatype} column,
that can only assume a finite number of values and specifies
the type of data stored in the specific entry.
With this schema, in exchange of a few bytes of extra information
for each attribute,
queries become efficient (for instance
by indexing the columns we can easily ask the database
backend to find all nodes that have a property with, e.g.,
\texttt{key}=``\texttt{energy}'', and \texttt{value}\,$\leq -12.3$).
Most importantly, we
achieved the main goal of being able to store \emph{any} value
in the database in a dynamical way (i.e., we do not have to
adapt the schema for each new attribute that we want to store).
Beside strings and numbers, we extended the EAV table to store also 
date/times and the datatype \texttt{None}.

An important extension is the implementation
of the storage of lists and dictionaries. These two datatypes are
common in the description of data inputs or outputs of atomistic simulations
(the list of atoms in a crystal structure, the $3\times 3$
components of the lattice vectors, \ldots).
One solution (implemented in AiiDA, but not used
by default) is to serialize any list or dictionary in the 
widespread JavaScript Object Notation (JSON) format as
a string, store the resulting string in the appropriate
column of \texttt{DbAttribute}, and set
the datatype field to a \texttt{JSON} value.
However, the disadvantage of this approach is that it becomes
inefficient to look for a specific value inside
a list: the values are now stored as strings, and a query
that looks for a given value must loop through 
the whole dataset (a task with $O(n)$ complexity and, moreover,
with a quite large prefactor because each string must be
deserialized -- i.e., the JSON string needs to be decoded).
Moreover, date/times are not natively supported by JSON.
We instead defined a schema that is able to store lists and dictionaries
at any depth (i.e. lists of lists, lists of dictionaries,
dictionaries of lists, \ldots).
We first add two new datatypes for lists and dictionaries, 
and we choose a separator symbol that is reserved and
cannot be used in key names (in the current implementation, a dot).
Then, for a dictionary with name \texttt{the\_dict} and values
$$
\mathtt{\{``value1": 12, ``value2": ``a\_string"\}}
$$
we store three entries in the \texttt{DbAttribute} table:
\begin{center}
\begin{tabular}{cccc}
key & datatype & int\_value & str\_value \\
\hline
\texttt{the\_dict} & \texttt{dict} & 2 & / \\
\texttt{the\_dict.value1} & \texttt{int} & 12 & / \\
\texttt{the\_dict.value2} & \texttt{str} & / & ``a\_string" 
\end{tabular}
\end{center}
The entry of type \texttt{dict} is needed to have the possibility to store
also empty dictionaries; the integer value is the number of elements
in the dictionary,
and is useful both for consistency checks and to be able to efficiently filter
dictionaries of a given length. The key of the other entries are obtained joining
the dictionary name, the separator (a dot) and the
key of the item.
Lists are stored in a similar way, where \texttt{value1} and
\texttt{value2} are replaced by the integer positions in the list,
and of course the datatype is replaced from \texttt{dict} to \texttt{list}.

We emphasize here that this storage model for lists and dictionaries
clearly extends naturally to any depth level (that is, in the example
above \texttt{the\_dict.value2} could in turn be a dictionary, 
and its values would have entries with key \texttt{the\_dict.value2.key1},
\texttt{the\_dict.value2.key2}, etc.)
Finally, we note here that while this storage model is efficient for 
storing and querying small lists, should not be used
for large arrays.
In this case, a specific class (\texttt{ArrayData}, see \ref{sec:plugins})
is available in the AiiDA ORM for storing directly the array on the
disk, rather than in the database. 

\subsection{\label{sec:transitiveclosure}Algorithm for the update of the transitive closure table}
Each entry of the transitive closure table 
represents the connectivity of two nodes in the
graph by means of a ``path'', i.e., a sequence of
links. 

This table is automatically updated by database triggers, that need to be fired every time an entry
is added, removed or modified in the table of 
links (\texttt{DbLink}).

The algorithm that we implemented is the following.
Every time a new link from a node $\mathcal{A}$ to a node $\mathcal{B}$
is created, the transitive closure table is queried to get all the
parents of $\mathcal{A}$ and the children of $\mathcal{B}$. Then, new paths are added, connecting: 
$\mathcal{A}$ to $\mathcal{B}$; each parent of $\mathcal{A}$
to $\mathcal{B}$; $\mathcal{A}$ to each child of $\mathcal{B}$; and each parent
of $\mathcal{A}$ to each child of $\mathcal{B}$.
Beside storing the PKs of the two nodes $\mathcal{A}$ and $\mathcal{B}$,
each entry of the \texttt{DbPath} table has three more columns to store
the PKs of the three \texttt{DbPath} entries that were used in the
creation of the path
(the one from the parent of $\mathcal{A}$ to $\mathcal{A}$, the
one from $\mathcal{A}$ to $\mathcal{B}$, and the one
from $\mathcal{B}$ to its child; the first and the third are set to
the PK of the $\mathcal{A}\to\mathcal{B}$ path, if absent).
These three additional columns
guarantee the possibility of implementing
an efficient algorithm to update the transitive closure table not only on
creation of new links, but also in the case of link removal: without 
them, the whole table would need to be regenerated from
scratch at every deletion.

\section*{Bibliography}
\let\oldthebibliography=\thebibliography
  \let\endoldthebibliography=\endthebibliography
  \renewenvironment{thebibliography}[1]{%
    \begin{oldthebibliography}{#1}%
      \setlength{\parskip}{0ex}%
      \setlength{\itemsep}{0ex}%
  }%
  {%
    \end{oldthebibliography}%
  }
\small\setlength{\parskip}{0ex}\setlength{\itemsep}{0ex}


\begin{thebibliography}{69}
\expandafter\ifx\csname natexlab\endcsname\relax\def\natexlab#1{#1}\fi
\providecommand{\bibinfo}[2]{#2}
\ifx\xfnm\relax \def\xfnm[#1]{\unskip,\space#1}\fi
\bibitem[{Van~Noorden et~al.(2014)Van~Noorden, Maher, and
  Nuzzo}]{VanNoorden:2014}
\bibinfo{author}{R.~Van~Noorden}, \bibinfo{author}{B.~Maher},
  \bibinfo{author}{R.~Nuzzo}, \bibinfo{journal}{Nature} \bibinfo{volume}{514}
  (\bibinfo{year}{2014}) \bibinfo{pages}{550}.
\bibitem[{Franceschetti and Zunger(1999)}]{Franceschetti:1999}
\bibinfo{author}{A.~Franceschetti}, \bibinfo{author}{A.~Zunger},
  \bibinfo{journal}{Nature} \bibinfo{volume}{402} (\bibinfo{year}{1999})
  \bibinfo{pages}{60--63}.
\bibitem[{J\'ohannesson et~al.(2002)J\'ohannesson, Bligaard, Ruban, Skriver,
  Jacobsen, and N\o{}rskov}]{Johannesson:2002}
\bibinfo{author}{G.~H. J\'ohannesson}, \bibinfo{author}{T.~Bligaard},
  \bibinfo{author}{A.~V. Ruban}, \bibinfo{author}{H.~L. Skriver},
  \bibinfo{author}{K.~W. Jacobsen}, \bibinfo{author}{J.~K. N\o{}rskov},
  \bibinfo{journal}{Phys. Rev. Lett.} \bibinfo{volume}{88}
  (\bibinfo{year}{2002}) \bibinfo{pages}{255506--}.
\bibitem[{Curtarolo et~al.(2003)Curtarolo, Morgan, Persson, Rodgers, and
  Ceder}]{Curtarolo:2003}
\bibinfo{author}{S.~Curtarolo}, \bibinfo{author}{D.~Morgan},
  \bibinfo{author}{K.~Persson}, \bibinfo{author}{J.~Rodgers},
  \bibinfo{author}{G.~Ceder}, \bibinfo{journal}{Phys. Rev. Lett.}
  \bibinfo{volume}{91} (\bibinfo{year}{2003}) \bibinfo{pages}{135503--}.
\bibitem[{Curtarolo et~al.(2013)Curtarolo, Hart, Nardelli, Mingo, Sanvito, and
  Levy}]{Curtarolo:2013}
\bibinfo{author}{S.~Curtarolo}, \bibinfo{author}{G.~L.~W. Hart},
  \bibinfo{author}{M.~B. Nardelli}, \bibinfo{author}{N.~Mingo},
  \bibinfo{author}{S.~Sanvito}, \bibinfo{author}{O.~Levy},
  \bibinfo{journal}{Nat. Mater.} \bibinfo{volume}{12} (\bibinfo{year}{2013})
  \bibinfo{pages}{191--201}.
\bibitem[{Villars et~al.(2004)Villars, Berndt, Brandenburg, Cenzual, Daams,
  Hulliger, Massalski, Okamoto, Osaki, Prince, Putz, and Iwata}]{Villars:2004}
\bibinfo{author}{P.~Villars}, \bibinfo{author}{M.~Berndt},
  \bibinfo{author}{K.~Brandenburg}, \bibinfo{author}{K.~Cenzual},
  \bibinfo{author}{J.~Daams}, \bibinfo{author}{F.~Hulliger},
  \bibinfo{author}{T.~Massalski}, \bibinfo{author}{H.~Okamoto},
  \bibinfo{author}{K.~Osaki}, \bibinfo{author}{A.~Prince},
  \bibinfo{author}{H.~Putz}, \bibinfo{author}{S.~Iwata},
  \bibinfo{journal}{Journal of Alloys and Compounds} \bibinfo{volume}{367}
  (\bibinfo{year}{2004}) \bibinfo{pages}{293--297}.
\bibitem[{Zarkevich(2006)}]{Zarkevich:2006}
\bibinfo{author}{N.~Zarkevich}, \bibinfo{journal}{Complexity}
  \bibinfo{volume}{11} (\bibinfo{year}{2006}) \bibinfo{pages}{36--42}.
\bibitem[{da~Silveira et~al.(2008)da~Silveira, da~Silva, and
  Wentzcovitch}]{Silveira:2008}
\bibinfo{author}{P.~R. da~Silveira}, \bibinfo{author}{C.~R. da~Silva},
  \bibinfo{author}{R.~M. Wentzcovitch}, \bibinfo{journal}{Comp. Phys. Comm.}
  \bibinfo{volume}{178} (\bibinfo{year}{2008}) \bibinfo{pages}{186--198}.
\bibitem[{Yuan and Gygi(2010)}]{Yuan:2010}
\bibinfo{author}{G.~Yuan}, \bibinfo{author}{F.~Gygi},
  \bibinfo{journal}{Computational Science \& Discovery} \bibinfo{volume}{3}
  (\bibinfo{year}{2010}) \bibinfo{pages}{015004--}.
\bibitem[{Jain et~al.(2011)Jain, Hautier, Moore, Ping~Ong, Fischer, Mueller,
  Persson, and Ceder}]{Jain:2011}
\bibinfo{author}{A.~Jain}, \bibinfo{author}{G.~Hautier}, \bibinfo{author}{C.~J.
  Moore}, \bibinfo{author}{S.~Ping~Ong}, \bibinfo{author}{C.~C. Fischer},
  \bibinfo{author}{T.~Mueller}, \bibinfo{author}{K.~A. Persson},
  \bibinfo{author}{G.~Ceder}, \bibinfo{journal}{Comp. Mat. Sci.}
  \bibinfo{volume}{50} (\bibinfo{year}{2011}) \bibinfo{pages}{2295--2310}.
\bibitem[{Adams et~al.(2011)Adams, de~Castro, Echenique, Estrada, Hanwell,
  Murray-Rust, Sherwood, Thomas, and Townsend}]{Adams:2011}
\bibinfo{author}{S.~Adams}, \bibinfo{author}{P.~de~Castro},
  \bibinfo{author}{P.~Echenique}, \bibinfo{author}{J.~Estrada},
  \bibinfo{author}{M.~D. Hanwell}, \bibinfo{author}{P.~Murray-Rust},
  \bibinfo{author}{P.~Sherwood}, \bibinfo{author}{J.~Thomas},
  \bibinfo{author}{J.~Townsend}, \bibinfo{journal}{Journal of Cheminformatics}
  \bibinfo{volume}{3} (\bibinfo{year}{2011}) \bibinfo{pages}{38}.
\bibitem[{Curtarolo et~al.(2012)Curtarolo, Setyawan, Wang, Xue, Yang, Taylor,
  Nelson, Hart, Sanvito, Buongiorno-Nardelli, Mingo, and Levy}]{Curtarolo:2012}
\bibinfo{author}{S.~Curtarolo}, \bibinfo{author}{W.~Setyawan},
  \bibinfo{author}{S.~Wang}, \bibinfo{author}{J.~Xue},
  \bibinfo{author}{K.~Yang}, \bibinfo{author}{R.~H. Taylor},
  \bibinfo{author}{L.~J. Nelson}, \bibinfo{author}{G.~L. Hart},
  \bibinfo{author}{S.~Sanvito}, \bibinfo{author}{M.~Buongiorno-Nardelli},
  \bibinfo{author}{N.~Mingo}, \bibinfo{author}{O.~Levy},
  \bibinfo{journal}{Computational Materials Science} \bibinfo{volume}{58}
  (\bibinfo{year}{2012}) \bibinfo{pages}{227--235}.
\bibitem[{Landis et~al.(2012)Landis, Hummelsh\o{}j, Nestorov, Greeley,
  Du\l{}ak, Bligaard, N\o{}rskov, and Jacobsen}]{Landis:2012}
\bibinfo{author}{D.~Landis}, \bibinfo{author}{J.~Hummelsh\o{}j},
  \bibinfo{author}{S.~Nestorov}, \bibinfo{author}{J.~Greeley},
  \bibinfo{author}{M.~Du\l{}ak}, \bibinfo{author}{T.~Bligaard},
  \bibinfo{author}{J.~N\o{}rskov}, \bibinfo{author}{K.~Jacobsen},
  \bibinfo{journal}{Computing in Science \& Engineering} \bibinfo{volume}{14}
  (\bibinfo{year}{2012}) \bibinfo{pages}{51--57}.
\bibitem[{Saal et~al.(2013)Saal, Kirklin, Aykol, Meredig, and
  Wolverton}]{Saal:2013}
\bibinfo{author}{J.~Saal}, \bibinfo{author}{S.~Kirklin},
  \bibinfo{author}{M.~Aykol}, \bibinfo{author}{B.~Meredig},
  \bibinfo{author}{C.~Wolverton} \bibinfo{volume}{65} (\bibinfo{year}{2013})
  \bibinfo{pages}{1501--1509}.
\bibitem[{eud(????)}]{eudat}
\bibinfo{howpublished}{http://www.eudat.eu/}.
\bibitem[{nom(????)}]{nomad}
\bibinfo{howpublished}{http://www.nomad-repository.eu/}.
\bibitem[{doc(????)}]{docker}
\bibinfo{howpublished}{https://www.docker.com/}.
\bibitem[{dja(????)}]{django}
\bibinfo{howpublished}{http://www.djangoproject.com/}.
\bibitem[{sql(????)}]{sqlalchemy}
\bibinfo{howpublished}{http://www.sqlalchemy.org/}.
\bibitem[{rub(????)}]{rubyonrails}
\bibinfo{howpublished}{http://rubyonrails.org/}.
\bibitem[{Bahn and Jacobsen(2002)}]{Bahn:2002}
\bibinfo{author}{S.~R. Bahn}, \bibinfo{author}{K.~W. Jacobsen},
  \bibinfo{journal}{Comput. Sci. Eng.} \bibinfo{volume}{4}
  (\bibinfo{year}{2002}) \bibinfo{pages}{56--66}.
\bibitem[{Ong et~al.(2013)Ong, Richards, Jain, Hautier, Kocher, Cholia, Gunter,
  Chevrier, Persson, and Ceder}]{Ong:2013}
\bibinfo{author}{S.~P. Ong}, \bibinfo{author}{W.~D. Richards},
  \bibinfo{author}{A.~Jain}, \bibinfo{author}{G.~Hautier},
  \bibinfo{author}{M.~Kocher}, \bibinfo{author}{S.~Cholia},
  \bibinfo{author}{D.~Gunter}, \bibinfo{author}{V.~L. Chevrier},
  \bibinfo{author}{K.~A. Persson}, \bibinfo{author}{G.~Ceder},
  \bibinfo{journal}{Comp. Mat. Sci.} \bibinfo{volume}{68}
  (\bibinfo{year}{2013}) \bibinfo{pages}{314--319}.
\bibitem[{Murray-Rust and Rzepa(2003)}]{Murray-Rust:2003}
\bibinfo{author}{P.~Murray-Rust}, \bibinfo{author}{H.~S. Rzepa},
  \bibinfo{journal}{J. Chem. Inf. Comput. Sci.} \bibinfo{volume}{43}
  (\bibinfo{year}{2003}) \bibinfo{pages}{757--772}.
\bibitem[{tco(????)}]{tcod}
\bibinfo{howpublished}{http://www.crystallography.net/tcod/}.
\bibitem[{cel(????)}]{celery}
\bibinfo{howpublished}{http://www.celeryproject.org/}.
\bibitem[{sup(????)}]{supervisor}
\bibinfo{howpublished}{http://supervisord.org/}.
\bibitem[{ope(????)}]{opengridforum}
\bibinfo{howpublished}{http://www.ogf.org}.
\bibitem[{Tr\"oger et~al.(2007)Tr\"oger, Rajic, Haas, , and
  Domagalski}]{Troger:2007}
\bibinfo{author}{P.~Tr\"oger}, \bibinfo{author}{H.~Rajic},
  \bibinfo{author}{A.~Haas}, , \bibinfo{author}{P.~Domagalski}, in:
  \bibinfo{booktitle}{Proceedings of the Seventh IEEE International Symposium
  on Cluster Computing and the Grid (CCGrid 2007), Rio de Janeiro, Brazil}, pp.
  \bibinfo{pages}{619--626}. \bibinfo{note}{{http://www.drmaa.org/}}.
\bibitem[{uni(????)}]{unicore}
\bibinfo{howpublished}{http://www.unicore.eu/}.
\bibitem[{gc3(????)}]{gc3pie}
\bibinfo{howpublished}{http://gc3pie.googlecode.com/}.
\bibitem[{tor(????)}]{torque}
\bibinfo{howpublished}{http://www.adaptivecomputing.com/products/open-source/torque/}.
\bibitem[{pbs(????)}]{pbspro}
\bibinfo{howpublished}{http://www.pbsworks.com/Product.aspx?id=1}.
\bibitem[{slu(????)}]{slurm}
\bibinfo{howpublished}{https://computing.llnl.gov/linux/slurm/}.
\bibitem[{ope(????)}]{opengridscheduler}
\bibinfo{title}{http://gridscheduler.sourceforge.net/}.
\bibitem[{par(????)}]{paramiko}
\bibinfo{howpublished}{https://github.com/paramiko/paramiko}.
\bibitem[{Moreau et~al.(2011)Moreau, Clifford, Freire, Futrelle, Gil, Groth,
  Kwasnikowska, Miles, Missier, Myers, Plale, Simmhan, Stephan, and den
  Bussche}]{Moreau:2011}
\bibinfo{author}{L.~Moreau}, \bibinfo{author}{B.~Clifford},
  \bibinfo{author}{J.~Freire}, \bibinfo{author}{J.~Futrelle},
  \bibinfo{author}{Y.~Gil}, \bibinfo{author}{P.~Groth},
  \bibinfo{author}{N.~Kwasnikowska}, \bibinfo{author}{S.~Miles},
  \bibinfo{author}{P.~Missier}, \bibinfo{author}{J.~Myers},
  \bibinfo{author}{B.~Plale}, \bibinfo{author}{Y.~Simmhan},
  \bibinfo{author}{E.~Stephan}, \bibinfo{author}{J.~V. den Bussche},
  \bibinfo{journal}{Future Generation Computer Systems} \bibinfo{volume}{27}
  (\bibinfo{year}{2011}) \bibinfo{pages}{743--756}.
\bibitem[{mys(????)}]{mysql}
\bibinfo{howpublished}{http://www.mysql.com/}.
\bibitem[{pos(????)}]{postgresql}
\bibinfo{howpublished}{http://www.postgresql.org/}.
\bibitem[{sql(????)}]{sqlite}
\bibinfo{howpublished}{http://www.sqlite.org/}.
\bibitem[{neo(????)}]{neo4j}
\bibinfo{howpublished}{http://www.neo4j.org/}.
\bibitem[{Dong et~al.(1999)Dong, Libkin, Su, and Wong}]{Dong:1999}
\bibinfo{author}{G.~Dong}, \bibinfo{author}{L.~Libkin},
  \bibinfo{author}{J.~Su}, \bibinfo{author}{L.~Wong}, \bibinfo{journal}{Int.
  Journal of Information Technology} \bibinfo{volume}{5} (\bibinfo{year}{1999})
  \bibinfo{pages}{46--78}.
\bibitem[{spg(????)}]{spglib}
\bibinfo{howpublished}{http://spglib.sourceforge.net}.
\bibitem[{van~der Walt et~al.(2011)van~der Walt, Colbert, and
  Varoquaux}]{vanderWalt:2011}
\bibinfo{author}{S.~van~der Walt}, \bibinfo{author}{S.~C. Colbert},
  \bibinfo{author}{G.~Varoquaux}, \bibinfo{journal}{Comput. Sci. Eng.}
  \bibinfo{volume}{13} (\bibinfo{year}{2011}) \bibinfo{pages}{22}.
\bibitem[{Setyawan and Curtarolo(2010)}]{Setyawan:2010}
\bibinfo{author}{W.~Setyawan}, \bibinfo{author}{S.~Curtarolo},
  \bibinfo{journal}{Computational Materials Science} \bibinfo{volume}{49}
  (\bibinfo{year}{2010}) \bibinfo{pages}{299--312}.
\bibitem[{ics(????)}]{icsd}
\bibinfo{howpublished}{http://www.fiz-karlsruhe.com/icsd.html}.
\bibitem[{Gra\v{z}ulis et~al.(2012)Gra\v{z}ulis, Da\v{s}kevi\v{c}, Merkys,
  Chateigner, Lutterotti, Quir\'os, Serebryanaya, Moeck, Downs, and
  Le~Bail}]{grazulis:2012}
\bibinfo{author}{S.~Gra\v{z}ulis}, \bibinfo{author}{A.~Da\v{s}kevi\v{c}},
  \bibinfo{author}{A.~Merkys}, \bibinfo{author}{D.~Chateigner},
  \bibinfo{author}{L.~Lutterotti}, \bibinfo{author}{M.~Quir\'os},
  \bibinfo{author}{N.~R. Serebryanaya}, \bibinfo{author}{P.~Moeck},
  \bibinfo{author}{R.~T. Downs}, \bibinfo{author}{A.~Le~Bail},
  \bibinfo{journal}{Nucleic Acids Research} \bibinfo{volume}{40}
  (\bibinfo{year}{2012}) \bibinfo{pages}{D420--D427}.
\bibitem[{P\'erez and Granger(2007)}]{Perez:2007}
\bibinfo{author}{F.~P\'erez}, \bibinfo{author}{B.~E. Granger},
  \bibinfo{journal}{Computing in Science and Engineering} \bibinfo{volume}{9}
  (\bibinfo{year}{2007}) \bibinfo{pages}{21--29}.
\bibitem[{gal(????)}]{galaxy}
\bibinfo{howpublished}{http://galaxyproject.org}.
\bibitem[{kep(????)}]{kepler}
\bibinfo{howpublished}{https://kepler-project.org}.
\bibitem[{kni(????)}]{knime}
\bibinfo{howpublished}{http://www.knime.org}.
\bibitem[{vis(????)}]{vistrails}
\bibinfo{howpublished}{http://www.vistrails.org}.
\bibitem[{Wolstencroft et~al.(2013)Wolstencroft, Haines, Fellows, Williams,
  Withers, Owen, Soiland-Reyes, Dunlop, Nenadic, Fisher, Bhagat, Belhajjame,
  Bacall, Hardisty, Nieva de~la Hidalga, Balcazar~Vargas, Sufi, and
  Goble}]{Wolstencroft:2013}
\bibinfo{author}{K.~Wolstencroft}, \bibinfo{author}{R.~Haines},
  \bibinfo{author}{D.~Fellows}, \bibinfo{author}{A.~Williams},
  \bibinfo{author}{D.~Withers}, \bibinfo{author}{S.~Owen},
  \bibinfo{author}{S.~Soiland-Reyes}, \bibinfo{author}{I.~Dunlop},
  \bibinfo{author}{A.~Nenadic}, \bibinfo{author}{P.~Fisher},
  \bibinfo{author}{J.~Bhagat}, \bibinfo{author}{K.~Belhajjame},
  \bibinfo{author}{F.~Bacall}, \bibinfo{author}{A.~Hardisty},
  \bibinfo{author}{A.~Nieva de~la Hidalga}, \bibinfo{author}{M.~P.
  Balcazar~Vargas}, \bibinfo{author}{S.~Sufi}, \bibinfo{author}{C.~Goble},
  \bibinfo{journal}{Nucleic Acids Research} \bibinfo{volume}{41}
  (\bibinfo{year}{2013}) \bibinfo{pages}{W557--W561}.
\bibitem[{Bart\'ok et~al.(2010)Bart\'ok, Payne, Kondor, and
  Cs\'anyi}]{Bartok:2010}
\bibinfo{author}{A.~P. Bart\'ok}, \bibinfo{author}{M.~C. Payne},
  \bibinfo{author}{R.~Kondor}, \bibinfo{author}{G.~Cs\'anyi},
  \bibinfo{journal}{Phys. Rev. Lett.} \bibinfo{volume}{104}
  (\bibinfo{year}{2010}) \bibinfo{pages}{136403--}.
\bibitem[{Glass et~al.(2006)Glass, Oganov, and Hansen}]{Glass:2006}
\bibinfo{author}{C.~W. Glass}, \bibinfo{author}{A.~R. Oganov},
  \bibinfo{author}{N.~Hansen}, \bibinfo{journal}{Comp. Phys. Comm.}
  \bibinfo{volume}{175} (\bibinfo{year}{2006}) \bibinfo{pages}{713--720}.
\bibitem[{Pickard and Needs(2011)}]{Pickard:2011}
\bibinfo{author}{C.~J. Pickard}, \bibinfo{author}{R.~J. Needs},
  \bibinfo{journal}{J. Phys. Cond. Matt.} \bibinfo{volume}{23}
  (\bibinfo{year}{2011}) \bibinfo{pages}{053201--}.
\bibitem[{sph(????)}]{sphinx}
\bibinfo{howpublished}{http://sphinx-doc.org}.
\bibitem[{rfc(????)}]{rfc4122}
\bibinfo{howpublished}{http://www.ietf.org/rfc/rfc4122.txt}.
\bibitem[{mpo(????)}]{mpod}
\bibinfo{howpublished}{http://mpod.cimav.edu.mx/index/}.
\bibitem[{Giannozzi et~al.(2009)Giannozzi, Baroni, Bonini, Calandra, Car,
  Cavazzoni, Ceresoli, Chiarotti, Cococcioni, Dabo, Corso, de~Gironcoli,
  Fabris, Fratesi, Gebauer, Gerstmann, Gougoussis, Kokalj, Lazzeri,
  Martin-Samos, Marzari, Mauri, Mazzarello, Paolini, Pasquarello, Paulatto,
  Sbraccia, Scandolo, Sclauzero, Seitsonen, Smogunov, Umari, and
  Wentzcovitch}]{Giannozzi:2009}
\bibinfo{author}{P.~Giannozzi}, \bibinfo{author}{S.~Baroni},
  \bibinfo{author}{N.~Bonini}, \bibinfo{author}{M.~Calandra},
  \bibinfo{author}{R.~Car}, \bibinfo{author}{C.~Cavazzoni},
  \bibinfo{author}{D.~Ceresoli}, \bibinfo{author}{G.~L. Chiarotti},
  \bibinfo{author}{M.~Cococcioni}, \bibinfo{author}{I.~Dabo},
  \bibinfo{author}{A.~D. Corso}, \bibinfo{author}{S.~de~Gironcoli},
  \bibinfo{author}{S.~Fabris}, \bibinfo{author}{G.~Fratesi},
  \bibinfo{author}{R.~Gebauer}, \bibinfo{author}{U.~Gerstmann},
  \bibinfo{author}{C.~Gougoussis}, \bibinfo{author}{A.~Kokalj},
  \bibinfo{author}{M.~Lazzeri}, \bibinfo{author}{L.~Martin-Samos},
  \bibinfo{author}{N.~Marzari}, \bibinfo{author}{F.~Mauri},
  \bibinfo{author}{R.~Mazzarello}, \bibinfo{author}{S.~Paolini},
  \bibinfo{author}{A.~Pasquarello}, \bibinfo{author}{L.~Paulatto},
  \bibinfo{author}{C.~Sbraccia}, \bibinfo{author}{S.~Scandolo},
  \bibinfo{author}{G.~Sclauzero}, \bibinfo{author}{A.~P. Seitsonen},
  \bibinfo{author}{A.~Smogunov}, \bibinfo{author}{P.~Umari},
  \bibinfo{author}{R.~M. Wentzcovitch}, \bibinfo{journal}{J. Phys. Cond. Matt.}
  \bibinfo{volume}{21} (\bibinfo{year}{2009}) \bibinfo{pages}{395502--}.
\bibitem[{Gra\v{z}ulis et~al.(2015)Gra\v{z}ulis, Merkys, Vaitkus, and
  Okuli\v{c}-Kazarinas}]{Grazulis:2015}
\bibinfo{author}{S.~Gra\v{z}ulis}, \bibinfo{author}{A.~Merkys},
  \bibinfo{author}{A.~Vaitkus}, \bibinfo{author}{M.~Okuli\v{c}-Kazarinas},
  \bibinfo{journal}{J. Appl. Cryst.} \bibinfo{volume}{48}
  (\bibinfo{year}{2015}) \bibinfo{pages}{85--91}.
\bibitem[{Mortensen et~al.(2005)Mortensen, Hansen, and
  Jacobsen}]{Mortensen:2005}
\bibinfo{author}{J.~J. Mortensen}, \bibinfo{author}{L.~B. Hansen},
  \bibinfo{author}{K.~W. Jacobsen}, \bibinfo{journal}{Phys. Rev. B}
  \bibinfo{volume}{71} (\bibinfo{year}{2005}) \bibinfo{pages}{035109}.
\bibitem[{Kucukbenli et~al.(2014)Kucukbenli, Monni, Adetunji, Ge, Adebayo,
  Marzari, de~Gironcoli, and Dal~Corso}]{Kucukbenli:2014}
\bibinfo{author}{E.~Kucukbenli}, \bibinfo{author}{M.~Monni},
  \bibinfo{author}{B.~Adetunji}, \bibinfo{author}{X.~Ge},
  \bibinfo{author}{G.~Adebayo}, \bibinfo{author}{N.~Marzari},
  \bibinfo{author}{S.~de~Gironcoli}, \bibinfo{author}{A.~Dal~Corso},
  \bibinfo{journal}{arXiv:1404.3015v1}  (\bibinfo{year}{2014}).
\bibitem[{psl(????)}]{pslibrary}
\bibinfo{howpublished}{http://qe-forge.org/gf/project/pslibrary/}.
\bibitem[{Garrity et~al.(2014)Garrity, Bennett, Rabe, and
  Vanderbilt}]{Garrity:2014}
\bibinfo{author}{K.~F. Garrity}, \bibinfo{author}{J.~W. Bennett},
  \bibinfo{author}{K.~M. Rabe}, \bibinfo{author}{D.~Vanderbilt},
  \bibinfo{journal}{Computational Materials Science} \bibinfo{volume}{81}
  (\bibinfo{year}{2014}) \bibinfo{pages}{446--452}.
\bibitem[{Marzari et~al.(1999)Marzari, Vanderbilt, De~Vita, and
  Payne}]{Marzari:1999}
\bibinfo{author}{N.~Marzari}, \bibinfo{author}{D.~Vanderbilt},
  \bibinfo{author}{A.~De~Vita}, \bibinfo{author}{M.~C. Payne},
  \bibinfo{journal}{Phys. Rev. Lett.} \bibinfo{volume}{82}
  (\bibinfo{year}{1999}) \bibinfo{pages}{3296--3299}.
\bibitem[{Borghi et~al.(2014)Borghi, Ferretti, Nguyen, Dabo, and
  Marzari}]{Borghi:2014}
\bibinfo{author}{G.~Borghi}, \bibinfo{author}{A.~Ferretti},
  \bibinfo{author}{N.~L. Nguyen}, \bibinfo{author}{I.~Dabo},
  \bibinfo{author}{N.~Marzari}, \bibinfo{journal}{Phys. Rev. B}
  \bibinfo{volume}{90} (\bibinfo{year}{2014}) \bibinfo{pages}{075135}.
\bibitem[{cas(????)}]{cassandra}
\bibinfo{howpublished}{http://cassandra.apache.org/}.
\bibitem[{cou(????)}]{couchdb}
\bibinfo{howpublished}{http://couchdb.apache.org/}.
\bibitem[{mon(????)}]{mongodb}
\bibinfo{howpublished}{http://www.mongodb.org/}.

\end{thebibliography}
\end{document}